\documentclass[fleqn,usenatbib,useAMS]{mnras}

\usepackage{graphicx}
\usepackage{lscape,supertabular,longtable,pdflscape}
\usepackage{lastpage}
\usepackage{amsmath}
\usepackage{url}
\usepackage{soul}
\usepackage[pagewise]{lineno}
\usepackage[T1]{fontenc}

\usepackage{txfonts}
\usepackage[svgnames]{xcolor}
\newcommand{\msun}{\ensuremath{M_{\odot}}}   
\newcommand{\mini}{\ensuremath{M_{\rm ini}}}                         
\newcommand{\lsun}{\ensuremath{\mathit{L}_{\odot}}}                  
\newcommand{\llsun}{\ensuremath{\log_{10}(L/\lsun)}}

\newcommand{\mdot}{\ensuremath{\dot{M}}}                             
\newcommand{\teff}{\ensuremath{\mathit{T}_{\rm eff}}}                

\newcommand{\vini}{\ensuremath{V_{\rm ini}}}                     
\newcommand{\vcrit}{\ensuremath{V_{\rm crit}}}                   

\title[Grids of stellar models at super-solar metallicity] {Grids of stellar models with rotation VII: Models from 0.8 to 300\,\msun\ at super-solar metallicity ($Z = 0.020$)}

\author[N. Yusof et al.]{Norhasliza  Yusof$^{1}$\thanks{E-mail: norhaslizay@um.edu.my}, 
Raphael Hirschi$^{2,3}$,
Patrick Eggenberger$^{4}$,
Sylvia Ekstr\"om$^{4}$,
\newauthor
Cyril Georgy$^{4}$,
Yves Sibony$^{4}$,
Paul A. Crowther$^{5}$,
Georges Meynet$^{4}$,
Hasan Abu Kassim$^{1}$,
\newauthor
Wan Aishah Wan Harun$^{1}$, 
Andr\'e Maeder$^{4}$,
Jose H. Groh$^{6}$,
Eoin Farrell$^{6}$,
Laura Murphy$^{6}$
 \\
 $^{1}$Department of Physics, Faculty of Science, University of Malaya, 50603 Kuala Lumpur, Malaysia\\
 $^{2}$Astrophysics Group, Keele University, Keele, Staffordshire ST5 5BG, UK\\
 $^{3}$Institute for Physics and Mathematics of the Universe (WPI), University of Tokyo, 5-1-5 Kashiwanoha, Kashiwa 277-8583, Japan\\
 $^{4}$Department of Astronomy, University of Geneva, Chemin Pegasi 51, 1290 Versoix, Switzerland\\
 $^{5}$Department of Physics \& Astronomy, University of Sheffield, Hounsfield Road, Sheffield, S3 7RH, UK\\
 $^{6}$School of Physics, Trinity College Dublin, the University of Dublin, College Green, Dublin\\
 }

\date{Accepted XXX. Received YYY; in original form ZZZ}

\pubyear{2021}

\begin{document}
\label{firstpage}
\pagerange{\pageref{firstpage}--\pageref{lastpage}}
\maketitle

\begin{abstract}
We present a grid of stellar models at super-solar metallicity ($Z=0.020$) extending the previous grids of Geneva models at solar and sub-solar metallicities. A metallicity of $Z=0.020$ was chosen to match that of the inner Galactic disk. A modest increase of 43\% (=0.02/0.014) in metallicity compared to solar models means that the models evolve similarly to solar models but with slightly larger mass loss. Mass loss limits the final total masses of the super-solar models to 35\,\msun\ even for stars with initial masses much larger than 100\,\msun. Mass loss is strong enough in stars above 20\,\msun\ for rotating stars (25\,\msun\ for non-rotating stars) to remove the entire hydrogen-rich envelope. Our models thus predict SNII below 20\,\msun\ for rotating stars (25\,\msun\ for non-rotating stars) and SNIb (possibly SNIc) above that. We computed both isochrones and synthetic clusters to compare our super-solar models to the Westerlund 1 (Wd1) massive young cluster. A synthetic cluster combining rotating and non-rotating models with an age spread between $\log_{10}{\rm (age/yr)}= $ 6.7 and 7.0 is able to reproduce qualitatively the observed populations of WR, RSG and YSG stars in Wd1, in particular their simultaneous presence at \llsun\ = 5-5.5. 
The quantitative agreement is imperfect and we discuss the likely causes: synthetic cluster parameters, binary interactions, mass loss and their related uncertainties. In particular, mass loss in the cool part of the HRD plays a key role.  
\end{abstract}


\section{Introduction}\label{sec:intro}

Large homogeneous grids of stellar models facilitate the analysis and interpretation of a wide range of observations. They also enable us to study the dependence of stellar evolution on key parameters like mass, metallicity and rotation. 
There are several large published grids of evolutionary models 
covering various mass and metallicity ranges and including various input physics.
Examples include the grid from \citet{2017ApJ...838..161S} focused on low mass stars with solar-scaled composition and the grids of evolutionary models for rotating main-sequence stars with initial composition tailored to the Galaxy and Magellanic Clouds and including transport by magnetic fields \citep{Brott2011}.
The PARSEC database \citep{bressan2012parsec, chen2015parsec} covers a broad range of metallicities $(0.0001 \leq {Z} \leq 0.04)$ and  initial masses up to 350 $M_\odot$. This database adopted solar abundances from \cite{caffau2011solar}.   
The MIST database \citep{Dotter2016,Choi2016} adopted solar-scaled abundances from \cite{asplund2009chemical} with a mass range from $0.1$ to  $300$ $M_\odot$  and metallicities within $(-4.0 \leq \mathrm{[Z/H]} \leq 0.5)$.
Finally, the BaSTI database \citep{pietrinferni2004large,pietrinferni2006large, hidalgo2018updated} includes a solar-scaled composition grid with initial composition ranging from $[\mathrm{Fe/H}]=-3.20$ to $+0.45$ and initial masses up to 15 $M_\odot$ and a grid with $\alpha$-enhanced heavy element distribution  \citep{pietrinferni2021updated}.

Grids of single star models with and without rotation at 
$Z=0.014, 0.006,0.002, 0.0004, 0.0$, thus covering a wide 
range of metallicities from solar to primordial stars via 
the metallicities of the LMC, SMC and I Zw 18 
\citep{Ekstrom2012,Eggenberger2021,
Georgy2013b,Groh2019a,Murphy2021} have  been completed 
using the Geneva Stellar Evolution Code \citep[GENEC; see ][for 
details]{2008Ap&SS.316...43E}. This paper extends the GENEVA grids 
of models to super-solar metallicity. 
The grid of models starting with \citet{Ekstrom2012} is a major 
update of the previous generation of GENEVA grids published in the 
1990s \citep[e.\,g.]{Schaller1992,Meynet1994} and 
\citet{Ekstrom2012} describes the updates in input physics between 
the two grids. Two major updates are first the inclusion of 
rotation in the models and second an update of the solar 
composition following the work of \citet{Asplund2005}. The 
reference solar metallicity used in the present grid is $Z=0.014$ 
\citep[versus $Z=0.02$ used in][]{Schaller1992}. 
A metallicity of $Z=0.02$ for this super-solar metallicity grid 
was chosen to match that of the inner Milky Way, including the 
Galactic Centre itself. There is a well established metallicity 
gradient in the Galactic disk, with slope --0.03 to --0.07 dex/kpc 
\citep{Balser2011}, such that the representative metallicity 
at the end of the Galactic Bar will be $\sim$0.15 dex or 40\%higher than in the Solar neighbourhood \citep[][$\log$ O/H+12 = 
8.69]{2021A&A...653A.141A}, although there is some evidence for 
azimuthal variations \citep{Davies2009}. Although 
the Galactic Centre region has been observed for a long 
time,improvements in instrumentation has led to large numbers of 
massive stars available for quantitative study 
\citep{Liermann2009, Clark2018A&A}. While stellar evolution 
properties could be extrapolated from solar metallicity models, it 
is preferable to provide stellar models tailored to the higher 
metallicity of the inner Galaxy, which is the goal of this paper.

Whilst some published grids of models also use a metallicity 
$Z=0.02$ \citep[e.\,g.][]{Schaller1992,eldridgevink06,BPASS2018}, 
the present grid of models is super-solar so should not be 
compared to the $Z=0.02$ models that consider $Z=0.02$ as their 
solar metallicity \citep[e.\,g.][]{Schaller1992,BPASS2018}. 
Instead, they can be compared to published super-solar models 
\citep[e.\,g. the 
$Z=0.04$ of][]{Meynet1994}. The main reason for this is that mass 
loss is scaled using the ratio of the metallicity of the models 
relative to the reference solar composition considered. In this 
context, the present grid of model corresponds to [Fe/H]$=0.155$ 
(or a factor of 1.429$=0.02/0.014$). This being said, given the 
many changes in input physics between this and published grids of 
super-solar models \citep[e.\,g.][]{Meynet1994} and the fact that 
most super-solar grids of models use a value of $Z$ that is twice 
the solar value (versus only 1.429 in this grid), such comparisons 
offer limited insight. The grid of super-solar rotating models 
closest to the present grid is the [Fe/H]$=0.25$ grid of the MIST 
database \citep{Dotter2016,Choi2016} and we compare the present 
models to the MIST grid in Sect.\,\ref{sec:obs}.

The present grid of models is tailored for the 
inner Galactic disk, which contains several massive young star clusters. The best studied massive young star cluster in the inner Galactic disk is Westerlund~1 (Wd1), at a distance of $\sim$4 kpc 
\citep{beasor2021age}, while there are also several older massive 
clusters at the end of the Galactic Bar which are rich
in red supergiants \citep{Davies2009}. 
Within the Galactic Centre, at a distance of 8.2 kpc (GM SHOULD BE PC NOT KPC) \citep{GalCentre2019}, there are several young high mass ($\geq 10^{4} M_{\odot}$) clusters including the Arches, Quintuplet and Galactic Centre clusters, plus a rich massive star population within the Central Molecular Zone \citep{Clark2021}. We compare the present models to these clusters in Sect.\,\ref{sec:obs}.

The models presented in this paper will also be useful for extra-galactic studies of metal-rich (massive) galaxies undergoing high star-formation rates. Within the Local Group, the present day metallicity of M31 is  considered to be highly supersolar based on strong-line H\,{\sc ii} region calibrations \citep{Zaritsky1994}, such that
$Z = 0.03$ is commonly adopted. However, more recent direct H\,{\sc ii} determinations infer a central metallicity of $\log$ O/H + 12 = 8.7 to 8.9 \citep{Zurita2012}, with similar abundances from early-type stars in the inner disk \citep{Venn2000, Smartt2001}, such that $Z = 0.02$ is more suitable to M31. Stellar abundances as high as $\log$ O/H + 12 = 9.0  have been obtained \citep{Trundle2002}, potentially attributable to azimuthal variations.
Beyond the Local Group, there are known to be many high metallicity star-forming regions  \citep{Bresolin2005} the most metal-rich being $\log$ O/H + 12 = 8.9, 60\% higher than the Sun, according to standard nebular diagnostics. \citet{Bresolin2016} have highlighted still higher stellar abundances of $\log$ O/H + 12 = 9.0 close to the centre of M83, with both stellar and nebular diagnostics favouring slightly super-solar abundances within the inner disk.

This paper is structured as follows. A summary of physical ingredients is provided in Sect.~\ref{sec:ingredients},  results are presented in Sect.~\ref{sec:SSmodels}, comparisons with observations are provided in Sect.~\ref{sec:obs} with a discussion and conclusions drawn in Sect.~\ref{sec:conclusions}.

\section{Physical ingredients of the models}\label{sec:ingredients}
The physical ingredients of the present grid of models are the same as in the other papers in the series for consistency. These are described in detail in \citet{Ekstrom2012} (solar grid hereinafter) and we only summarize them here.

The initial composition of the models is given in Table\,\ref{tab:initial_abundances}. In particular, the initial abundances of H, He, and metals are set to ${\rm X} = 0.7064$, ${\rm Y} = 0.2735$, and ${\rm Z} = 0.02$. The mixture of heavy elements is solar-scaled \citep[scaled from $Z=0.014$ to $Z=0.02$ compared to][]{Ekstrom2012} with the solar mixture based on \citet{asp05} except for the Ne abundance, which is based on the work by \citet{cunha06}. Using this scaling, [Fe/H]$=0.155$. Isotopic ratios are taken from \citet{Lodders2003}.

\begin{table}
    \centering
    \caption{Initial composition of the models. The number in bracket is the exponent: e.\,g. $4.540\  (-5)=4.540\times 10^{-5}$.}
    \begin{tabular}{cc|cc}
     Nuclide    & Initial mass fraction & Nuclide    & Initial mass fraction \\
\hline    
 $^{1}$H      & 7.064 (-1)& $^{17}$O     & 3.237 (-6) \\
 $^{3}$He     & 4.540 (-5)& $^{18}$O     & 1.843 (-5)\\
 $^{4}$He     & 2.735 (-1)& $^{20}$Ne    & 2.681 (-3)\\
 $^{12}$C     & 3.261 (-3)& $^{22}$Ne    & 2.169 (-4)\\
 $^{13}$C     & 3.958 (-5)& $^{24}$Mg    & 7.193 (-4)\\
 $^{14}$N     & 9.411 (-4)& $^{25}$Mg    & 9.488 (-5)\\
 $^{15}$N     & 3.707 (-6)& $^{26}$Mg    & 1.086 (-4)\\
 $^{16}$O     & 8.169 (-3)& & 
    \end{tabular}
    \label{tab:initial_abundances}
\end{table}

The Schwarzschild criterion is used to determine the location of convective boundaries. Convective boundary mixing is only applied to hydrogen and helium burning cores in the form of overshooting with an overshooting distance $l_{\mathrm{ov}}=0.1\,H_P$  for $M \ge 1.7\, \msun$, 0.05 $H_\mathit{P}$ between 1.25 and 1.5 $\msun$, and 0 below ( where $H_\mathit{P}$ is the pressure scale-height scale at the Schwarzschild convective boundary). 
Studies such as \citet{Castro2014} observe a wider main sequence (MS) width for massive stars than predicted by models using $l_{\mathrm{ov}}=0.1\,H_P$. Models using a larger value of overshoot \citep[e.g. 0.035 in][]{Brott2011} predict a larger main sequence that fits the MS width infered for 15\,$\msun$ stars by \citet{Castro2014}  but still fail to explain the mass dependence of the MS width. The uncertainties linked to convective boundary mixing (CBM) and their impact on the evolution of massive stars have been studied extensively  \citep[see e.\,.g][]{Vink2010,Higgins2019,Davis2019,Kaiser2020, Martinet2021,Scott2021}. 
These studies generally find that using larger CBM (such as overshoot) 
leads to larger convective cores, higher luminosities and models behaving like more massive models with less CBM. Using larger CBM would for example tend to decrease the minimum mass for a single star to become a WR star. 
We nevertheless continue using $l_{\mathrm{ov}}=0.1\,H_P$ in this super-solar grid of models for consistency with the grids at other metallicities.

The stellar equations are modified to include the effects of rotation using the shellular-rotation hypothesis. The main rotation-induced instabilities included in the models are meridional 
circulation and (secular and dynamical) shear. For the transport of angular momentum, meridional circulation is implemented as an advective process {during the MS phase} while shear is implemented as a diffusive process\footnote{After the MS phase, in the present models, the main effect impacting the internal rotation is the local conservation of the angular momentum.}. Both processes are implemented in a diffusive approach for the transport of chemical elements \citep[see][for more details and references]{Ekstrom2012}. Magnetic instabilities are not included in the grids of models. 

The recipes for mass-loss rates ($\mdot$) used depend on mass, surface composition and position in the Hertzsprung-Russell (HR) diagram, and for consistency we follow the approach of previous grids.
On the MS, stars with a mass below 7 $M_{\sun}$ are computed at constant mass. Above 7 $M_{\sun}$, the radiative mass loss rate adopted is from \citet{vink01}. In the domains not covered by this prescription, the prescription from \citet{dJ88} is used.
For red (super)giants (RG/RSG), the \citet{Reimers75,Reimers77} formula (with $\eta=0.5$) is used for stars up to 12 $M_{\sun}$. The \citet{dJ88} prescription is applied from 15 $M_{\sun}$ and above for models with $\log (T_\text{eff}) > 3.7$. For $\log (T_\text{eff}) \leq 3.7$, a linear fit of the data from \citet{sylvester98} and \citet{vloon99} \citep[see][]{Crowther00} is used. 
Massive star models in the RSG phase sometimes have layers that exceed the Eddington luminosity limit. There are no theoretical prescriptions for mass loss in the RSG phase, and no precise observational or theoretical guidance for cases when the Eddington luminosity is exceeded. In order to nevertheless take into account when models exceed the Eddington limit, mass loss rates are increased by a factor of 3 whenever the luminosity of any of the layers of the envelope is higher than 5 times the Eddington luminosity \citep[see][for more details and a discussion on this topic]{Ekstrom2012}.
WR stars are computed with the \citet{nuglam00} prescription, or the\citet{grafham08} recipe in the small validity domain of this prescription. In some cases the WR mass loss rate from \citet{grafham08} is lower than the rate from \citet{vink01}. In these cases, the \citet{vink01} prescription is used instead. Both the \citet{nuglam00} 
and \citet{grafham08} mass loss rates account for some clumping effects 
\citep{muijres11} and are a factor of 2 to 3 smaller than the ``normal'' rates used in the 1992 grids \citep{Schaller1992}.

For rotating models, a correction factor is applied to the radiative mass loss rate as described in \citet{mm6}:
\begin{eqnarray}
\dot{M}(\Omega) &=& F_{\Omega}\cdot \dot{M}(\Omega=0)= F_{\Omega}\cdot \dot{M}_\text{rad}  \nonumber \\
& &\text{with} \hspace{.3cm}F_{\Omega}=\frac{(1-\Gamma)^{\frac{1}{\alpha}-1}}{\left[ 1-\frac{\Omega^2}{2\pi G \rho_\text{m}} - \Gamma \right]^{\frac{1}{\alpha}-1}}
\label{EqMdotRot}
\end{eqnarray}
where $\Gamma=L/L_\text{Edd}=\kappa L / (4\pi cGM)$ is the Eddington factor (with $\kappa$ is the electron-scattering opacity), $\Omega$ is the angular velocity and $\alpha$ the force multiplier parameter depending on $T_\text{eff}$.

Historically, empirical mass loss rates were derived using a 
mixture of rotating and non-rotating stars. To compensate for this 
fact, $\dot{M}(\Omega=0)$ is set to 0.85 times the mass loss rate 
obtained from the prescriptions above during the MS (main phase 
during which rotation rates are significant). This reduction factor does not need to be applied to the theoretical mass loss rate of 
\citet{vink01} but the 0.85 factor was still used in this grid of 
models for historical reasons and consistency with the grids at 
other metallicities.
For the same historical reasons some MESA models 
\citep[][]{Farmer2016,Ritter2018} apply a factor of 0.8 to mass 
loss prescriptions. We no longer recommend to use such reduction 
factor, especially for theoretical mass loss prescriptions such as 
\citet{vink01} or phases during which the average rotation rate is 
small (e.g. RSG phase).

The impact of the 0.85 reduction factor applied during the MS and 
in general of the mass loss enhancement factor due to rotation 
($F_{\Omega}$) remains very modest in the present grid of models 
and do not affect our conclusions.
Indeed, the rotating 20 and 25\,$\msun$ models lose 2.88 and 
0.63\,$\msun$ respectively (see Table\,\ref{TabListModels}) during 
the MS when these factors would modify the mass loss rate. This is 
much smaller than the mass loss in the RSG phase (more than 
10\,$\msun$ for both models), during which rotation is very slow 
and the mass loss rate applied are not modified by Eq.\,(1).
So most of the mass loss in the 20-25\,$\msun$ mass range is lost 
during RSG phase where rotating rates are low and the key factor 
determining mass loss is the luminosity. The other effects of 
rotation, rotation-induced mixing in particular, have a much larger impact on mass loss than the enhancement factors above by helping 
models in this mass range to reach the RSG early. 
For higher initial masses, $M \gtrsim 40\,\msun$, mass loss during 
the MS becomes significant (half of the initial mass or more for 
$M\gtrsim 85\,\msun$) so the enhancement factor above may play a 
role, especially if the model is close to the Eddington limit. In 
the present grid of models, however, mass loss is strong in both 
rotating and non-rotating model, keeping the very massive models 
away from the Eddington limit and the dominant impact of rotation 
is its indirect effects on the effective temperature and luminosity of the models.

Mass-loss rates are scaled with metallicity in the following way: $\mdot (Z) = (Z/Z_{\sun})^{\alpha} \mdot(Z_{\sun})$. For the MS and blue supergiant phases, we assume $\alpha=0.85$ or 0.50 when the 
\citet{vink01} or \citet{dJ88} recipes are used, respectively. For 
the Wolf-Rayet (WR) phase, we assume $\alpha=0.66$, following 
\citet{eldridgevink06}. For other phases, such as when the 
effective temperature $\teff$ is lower than $\log (\teff / K = 
3.7)$, no metallicity scaling is applied. Given the ratio of 
$0.02/0.014=1.43$, mass loss rates are larger by a factor between 1 ($\log_{10} (\teff / K) < 3.7$) and 1.35 ($\alpha=0.85$) in a 
super-solar model compared to the corresponding solar metallicity 
model.

\section{Properties of the stellar models}\label{sec:SSmodels}
We computed stellar evolution models for the following initial 
masses: 0.8, 0.9, 1, 1.1, 1.25, 1.35, 1.5, 1.7, 2, 2.5, 3, 4, 5, 7, 9, 12, 15, 20, 25, 32, 40, 60, 85, 120, 150, 200 and 300\,$\msun$. For each mass, we computed both a non-rotating and a rotating model with a ratio between the equatorial surface rotational velocity 
(\vini) and critical rotational velocity (\vcrit) of 0.4 (0 for the non-rotating models) at the zero-age main sequence (ZAMS). The 
models are evolved up to the end of core carbon burning ($\mini \ge 12\,\msun$), the early asymptotic giant branch ($2.5\,\msun \le 
\mini \le 9\,\msun$), or the helium flash ($\mini \le 2\,\msun$).

\begin{figure*}
    \centering
    \includegraphics[width=0.5\textwidth]{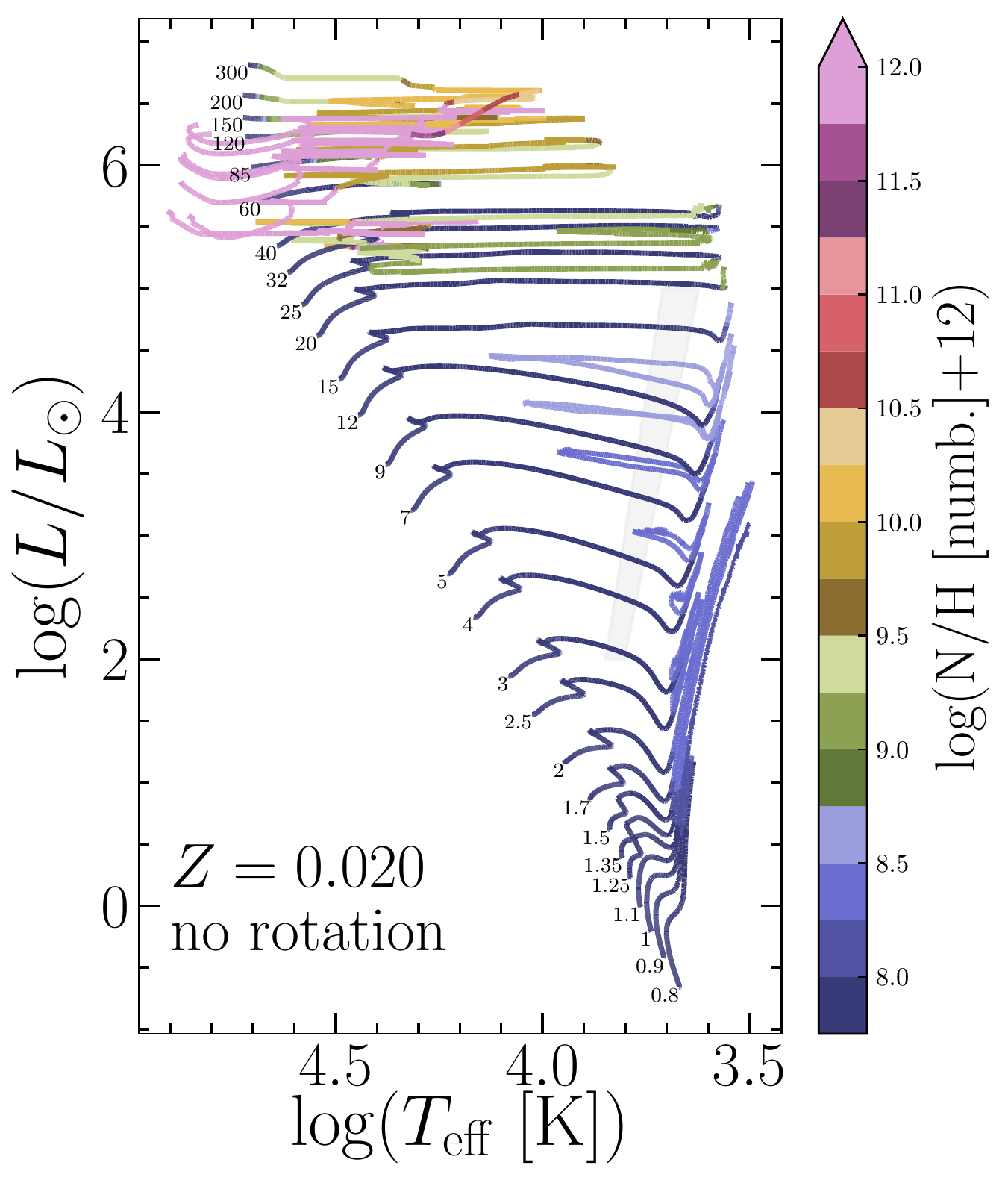}\includegraphics[width=0.5\textwidth]{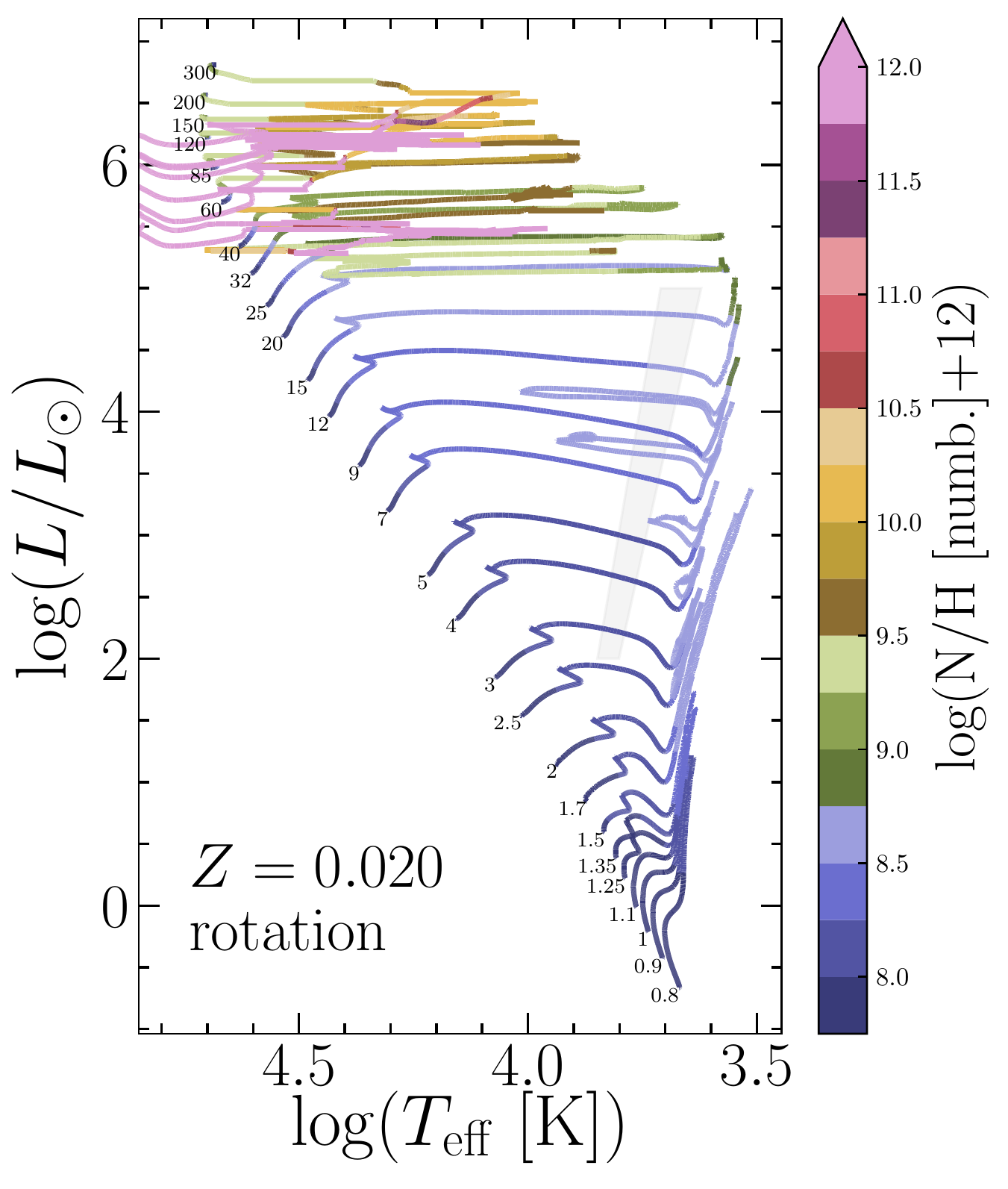}
    \caption{HR diagram for the non-rotating ({\it left}) and rotating ({\it right}) models at $Z=0.020$. Colour-coded is the surface nitrogen abundance in $\log(N/H)+12$. Note that the effective temperature used throughout the paper is the temperature at the surface of the star. The corrected temperature for thick winds is available in the electronic tables (see link at the end of the paper).}
    \label{fig:HRD_NH}
\end{figure*}

The main properties of the models at key stages (ZAMS, TAMS, and end of He- and C-burning phases if relevant) are presented in Tables \ref{Table:MfinMcores} and \ref{TabListModels}.
Similarly to \citet{Ekstrom2012} and \citet{Georgy2013b}, electronic tables of the evolutionary sequences are publicly available\footnote{See \href{http://obswww.unige.ch/Recherche/evol/-Database-}{http://obswww.unige.ch/Recherche/evol/-Database-}\\or the CDS database at \href{http://vizier.u-strasbg.fr/viz-bin/VizieR-2}{http://vizier.u-strasbg.fr/viz-bin/VizieR-2}.}. For each model, the evolutionary track is described by 400 selected data points, with each one corresponding to a given evolutionary stage. Points of different evolutionary tracks with the same number correspond to similar stages to facilitate the interpolation of evolutionary tracks. The points are numbered as described in \citet{Ekstrom2012}. The grids can thus be used as input for computing interpolated tracks, isochrones, and population synthesis models using the publicly available Geneva tools\footnote{\href{https://obswww.unige.ch/Recherche/evoldb/index/}{https://obswww.unige.ch/Recherche/evoldb/index/}}. A detailed description of the online tools is presented in \citet{Georgy2014}.

\subsection{Evolution of surface properties and lifetimes}\label{surfaceprop}
The evolution of the models in the Hertzsprung-Russell diagram (HRD) is presented in Fig.\,\ref{fig:HRD_NH} for the non-rotating ({\it left}) and rotating ({\it right}) models. 
In non-rotating models, the following features can be seen. The MS becomes significantly broader for stars above 30\,\msun\ due to the large convective cores and mass loss during H-burning \citep[see][for extended discussions on the MS width for massive stars and convective boundary mixing]{Vink2010,Castro2014,Higgins2019,Davis2019,Kaiser2020,Scott2021,Martinet2021}.
The strong mass loss in very massive stars (VMS, above 100\,\msun) leads to the tracks converging to the same luminosity range by the end of the MS.
The maximum luminosity of RSGs is around \llsun=5.7. Stars in the mass range between 25 and 40\,\msun\ evolve back to the blue side of the HRD after the RSG phase, while stars below this end their evolution as RSG/AGB/RG. Extended blue loops crossing the Cepheid instability strip occur in models between 5 and 12\,\msun.

\begin{figure*}
    \centering
    \includegraphics[width=.5\textwidth]{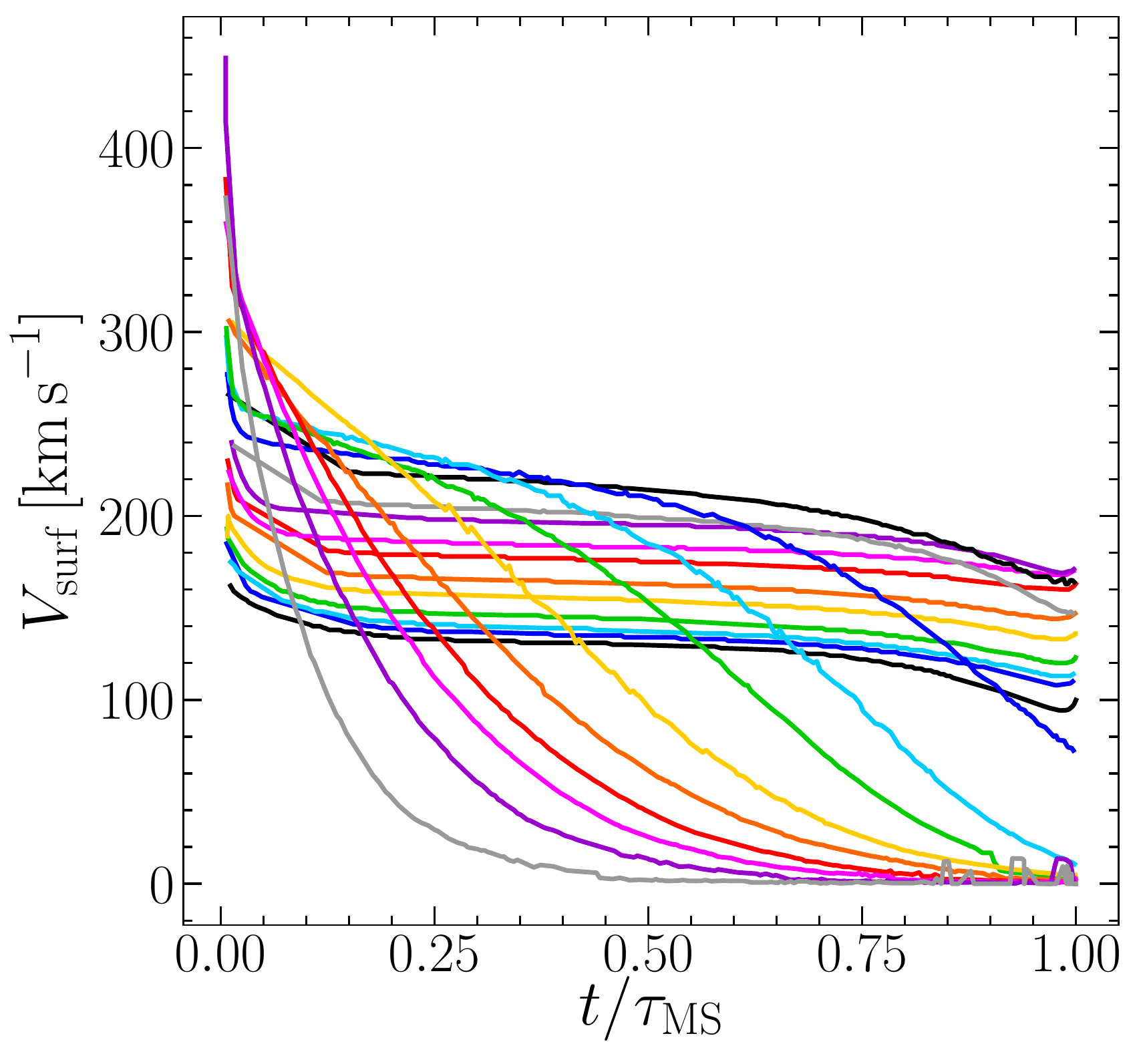}\includegraphics[width=.5\textwidth]{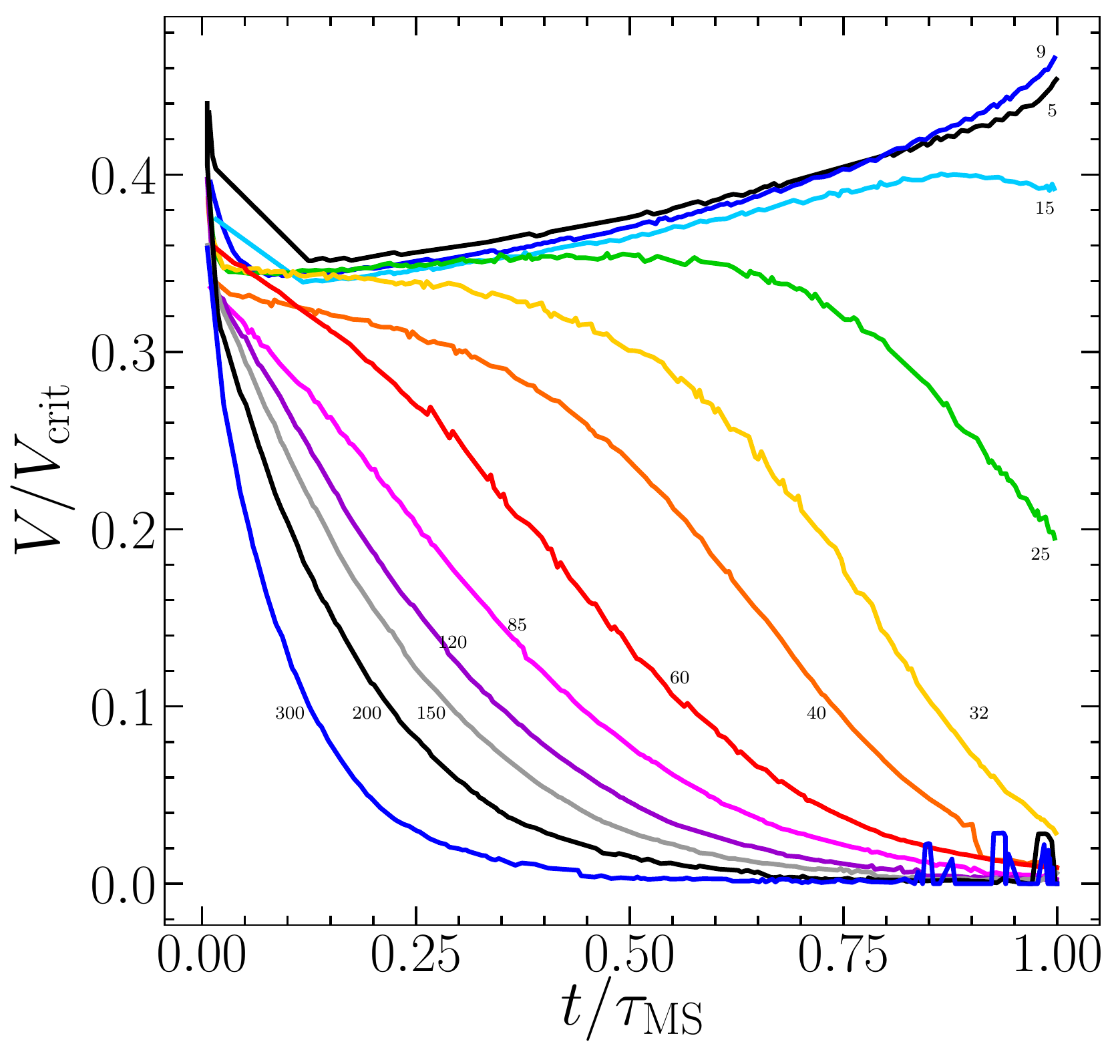}
    \caption{Velocity evolution of the rotating models at $Z=0.020$ as a function of the MS lifetime. {\it Left:} surface velocity. Stars from $1.7$ (lowest track at $t/\tau_{\mathrm{MS}}=0$) to $300\,\msun$. {\it Right:} ratio $V/V_{\mathrm{crit}}$ of selected massive star models.}
    \label{fig:rotEvol_Z020}
\end{figure*}

Rotation-induced mixing extends the MS lifetime (see Table\,\ref{TabListModels}) and luminosity of stars in general. Mixing of helium in the radiative zone above the core can make the MS width narrower \citep[see ][]{Martinet2021}, especially for stars with masses above 30\,\msun. The mixing of helium also generally tends to reduce the importance of the H-burning shell and rotating models reach the RSG earlier than non-rotating ones during He-burning \citep[their fig.\,3]{Hirschi2004}. All these effects (coupled with the reduced gravity discussed in the previous section) lead to stronger mass loss in rotating stars. This shifts the mass ranges mentioned above for non-rotating models to lower initial masses. The maximum luminosity of RSGs is \llsun=5.5 with stars around \llsun=5.6-5.8 being yellow super/hypergiants (YSG/YHG). Rotating stars from 20\,\msun\ upwards evolve back to the blue side of the HRD after the RSG phase. This means that stars with \llsun $\sim\ 5.2$ can occupy the full width of the HRD. This will be further discussed in Sect.\,\ref{sec:obs}.
Extended blue loops crossing the Cepheid instability strip occur in rotating models between 7 and 9\,\msun. The colour-coding for the nitrogen surface abundance shows that this enrichment occurs already during the MS in rotating models, while it only starts in the cool parts of the HRD for non-rotating stars below 50\,\msun.

Given the strong mass loss and related angular momentum loss experienced by massive stars at high metallicities, the surface rotation velocity of the models decreases during the MS and the rate of decrease increases with initial mass (see Fig.\,\ref{fig:rotEvol_Z020}, {\it left}). Massive stars above 15\,\msun\ thus move away from critical rotation (see Fig.\,\ref{fig:rotEvol_Z020}, {\it right}). The average surface rotation velocity of massive start on the MS is thus relatively low with $\bar{V}_\text{MS} \lesssim 200$\,km/s for $M_{\textrm{ini}} \ge 15$\,\msun (and $\bar{V}_\text{MS} \lesssim 100$\,km/s for $M_{\textrm{ini}} \ge 85$\,\msun).  In stars below  15\,\msun, internal transport of angular momentum leads models to get slightly closer to critical rotation.

\begin{figure*}
    \centering
    \includegraphics[width=0.5\textwidth]{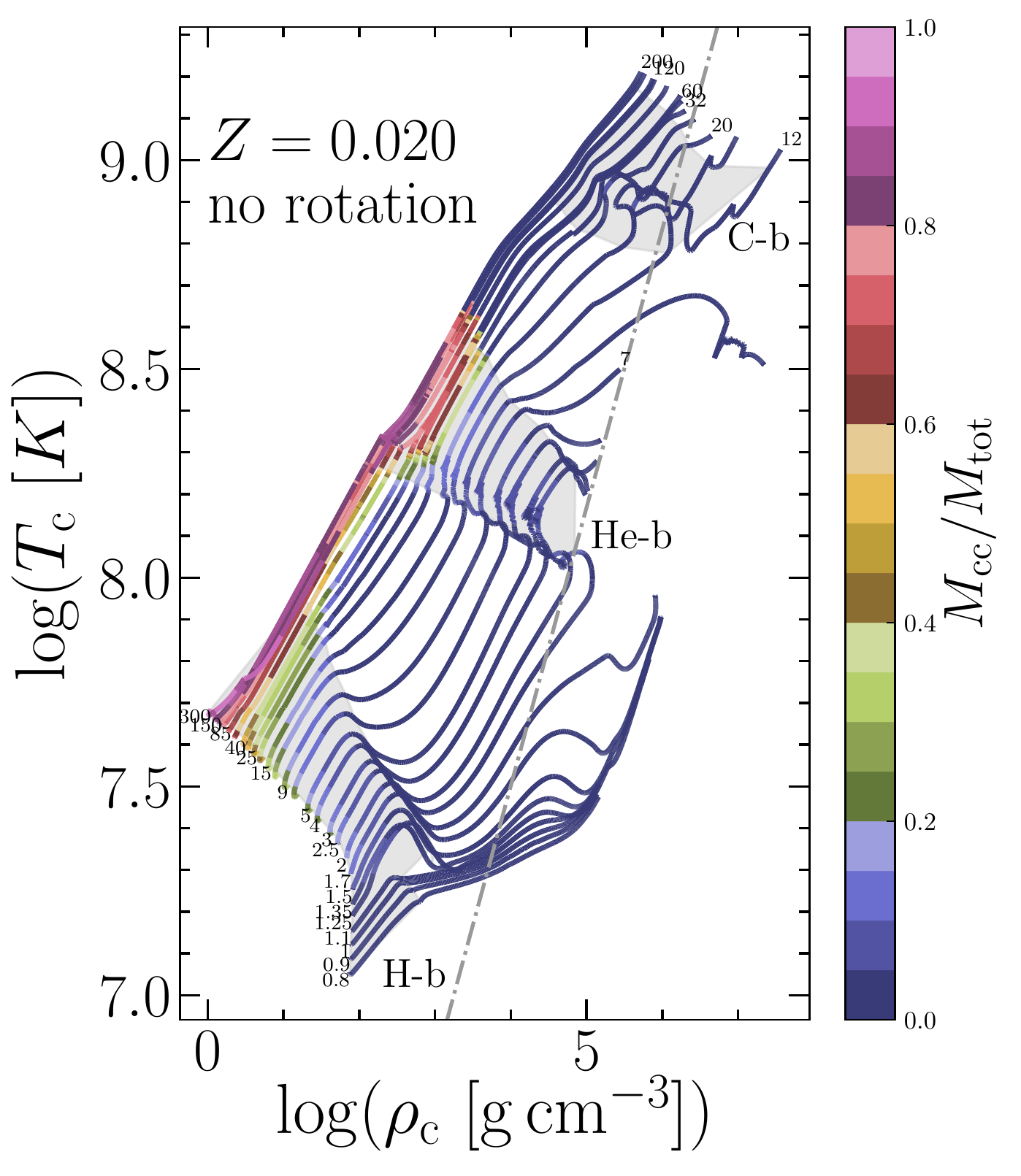}\includegraphics[width=0.5\textwidth]{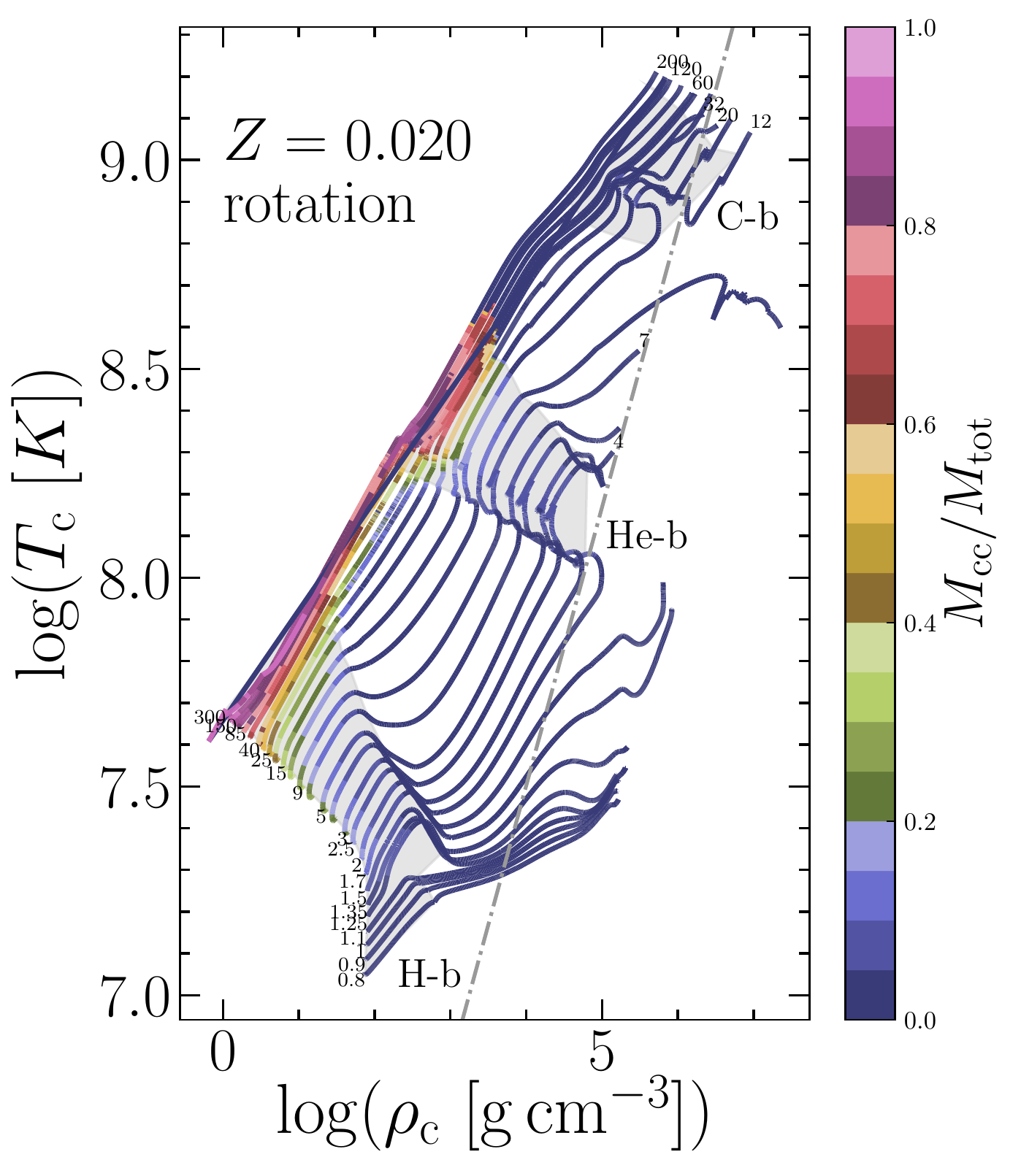}
    \caption{Central temperature ($T_\text{c}$) versus central density ($\rho_\text{c}$) diagram for the non-rotating ({\it left}) and rotating ({\it right}) models at $Z=0.020$. Colour-coded is the fractional mass ($M_\text{cc}/M_\text{tot}$) of the convective core. The dotted-dashed line indicates the transition from non-degenerate to partially-degenerate conditions.}
    \label{fig:rhoT_Mcc}
\end{figure*}

\begin{table}
    \centering
    \caption{Initial mass ($M_{\textrm{ini}}$) and ratio of the initial equatorial surface velocity ($V_\text{ini}/V_\text{crit}$) followed by the final total ($M_{\textrm{fin}}$), helium-core ($M_{\alpha,01}$ defined as the mass coordinate where the hydrogen mass fraction drops below 1\%) and carbon-oxygen core masses ($M_{\textrm{CO},01}$ defined as the mass coordinate where the helium mass fraction drops below 1\% and $M_{\textrm{CO},20}$ defined as the mass coordinate where the sum of the mass fractions of carbon and oxygen becomes larger than 20\%) of the models.}
    \begin{tabular}{cccccc}
     $M_{\text{ini}}$ & $V_\text{ini}/V_\text{crit}$ &$M_{\textrm{fin}}$ & $M_{\alpha,01}$ &$M_{\textrm{CO},01}$ & $M_{\textrm{CO},20}$ \\
\hline    
9 & 0.0 &   8.80 &   1.21 &   1.14 &   1.15 \\
9 & 0.4 &   8.74 &   1.83 &   1.31 &   1.48 \\
12 & 0.0 &  11.56 &   2.98 &   1.58 &   1.66 \\
12 & 0.4 &  10.36 &   3.68 &   2.14 &   3.06 \\
15 & 0.0 &  13.09 &   4.09 &   2.24 &   2.55 \\
15 & 0.4 &  10.83 &   5.22 &   3.09 &   4.86 \\
20 & 0.0 &   8.45 &   6.03 &   3.68 &   3.96 \\
20 & 0.4 &   7.27 &   7.14 &   4.66 &   7.09 \\
25 & 0.0 &   8.04 &   8.04 &   5.37 &   6.55 \\
25 & 0.4 &   9.08 &   9.08 &   6.67 &   8.95 \\
32 & 0.0 &  10.71 &  10.71 &   7.77 &   8.42 \\
32 & 0.4 &   9.80 &   9.80 &   7.16 &   9.80 \\
40 & 0.0 &  11.33 &  11.33 &   8.64 &  11.33 \\
40 & 0.4 &  11.63 &  11.63 &   8.97 &  11.63 \\
60 & 0.0 &  10.77 &  10.77 &   8.24 &  10.77 \\
60 & 0.4 &  12.87 &  12.87 &   9.93 &  12.87 \\
85 & 0.0 &  16.21 &  16.21 &  12.91 &  16.21 \\
85 & 0.4 &  16.64 &  16.64 &  13.25 &  16.64 \\
120 & 0.0 &  23.40 &  23.40 &  19.15 &  23.40 \\
120 & 0.4 &  22.26 &  22.26 &  18.05 &  22.26 \\
150 & 0.0 &  30.92 &  30.92 &  26.07 &  30.92 \\
150 & 0.4 &  25.79 &  25.79 &  21.00 &  25.79 \\
200 & 0.0 &  35.65 &  35.65 &  30.02 &  35.65 \\
200 & 0.4 &  34.64 &  34.64 &  29.09 &  34.64 \\
300 & 0.0 &  22.23 &  22.23 &  18.08 &  22.23 \\ 
300 & 0.4 &  25.24 &  25.24 &  20.62 &  25.24 
\end{tabular}
\label{Table:MfinMcores}
\end{table}

\subsection{Evolution of central properties and final total and core masses}
Rotation-induced mixing brings additional fuel into convective core and rotating models having generally larger central temperatures and lower central densities, thus behaving in their core like more massive non-rotating stars \citep[see][]{Hirschi2004}. This can be best seen in Fig.\,\ref{fig:rhoT_Mcc} by comparing the tracks of the 12\,\msun\ models in the partially degenerate section of the central temperature versus central density diagram. The convergence of the evolution tracks observed in the HRD for very massive stars is also observed in this diagram (for the same reason: strong mass loss).

\begin{figure*}
    \centering
    \includegraphics[width=.5\textwidth]{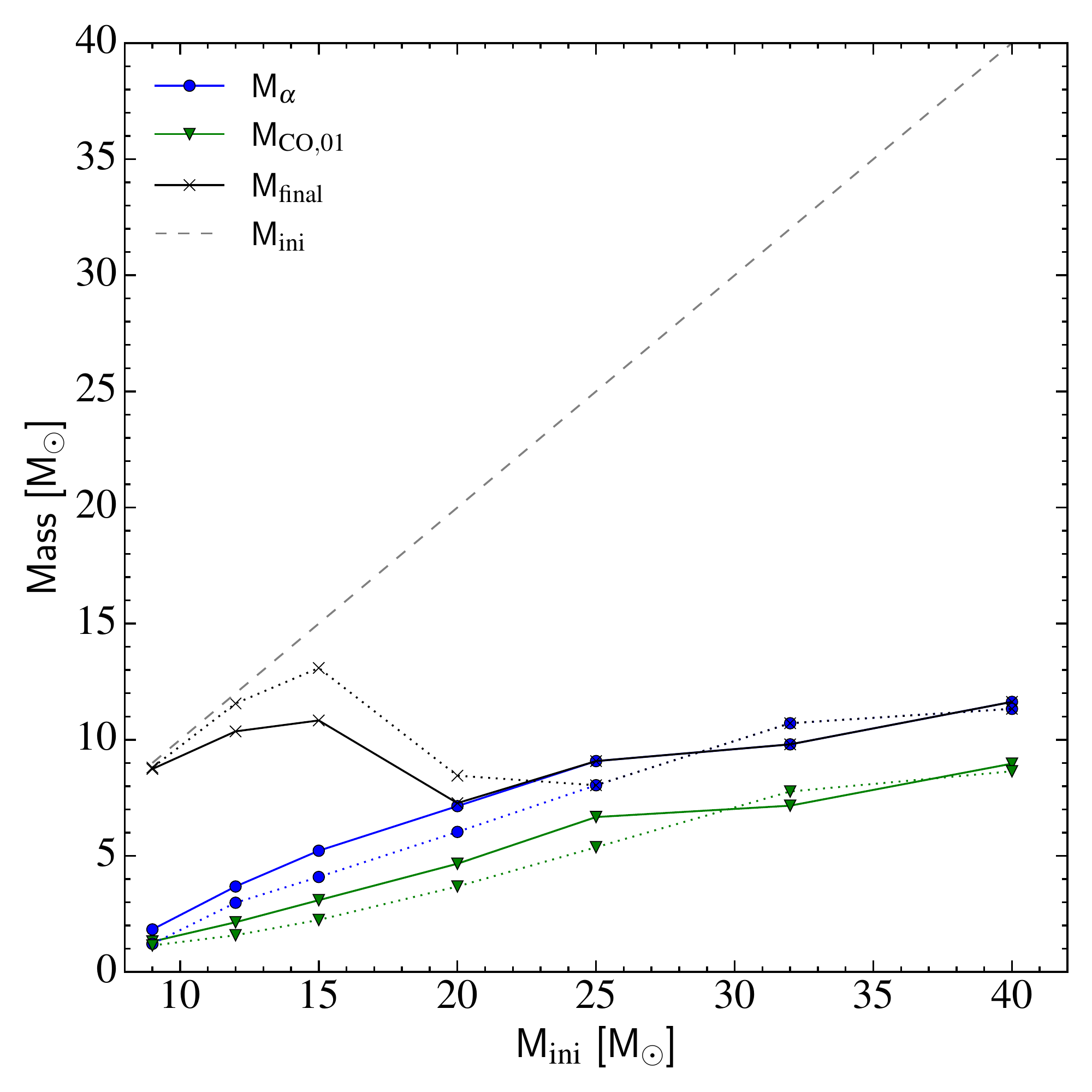}\hfill\includegraphics[width=.5\textwidth]{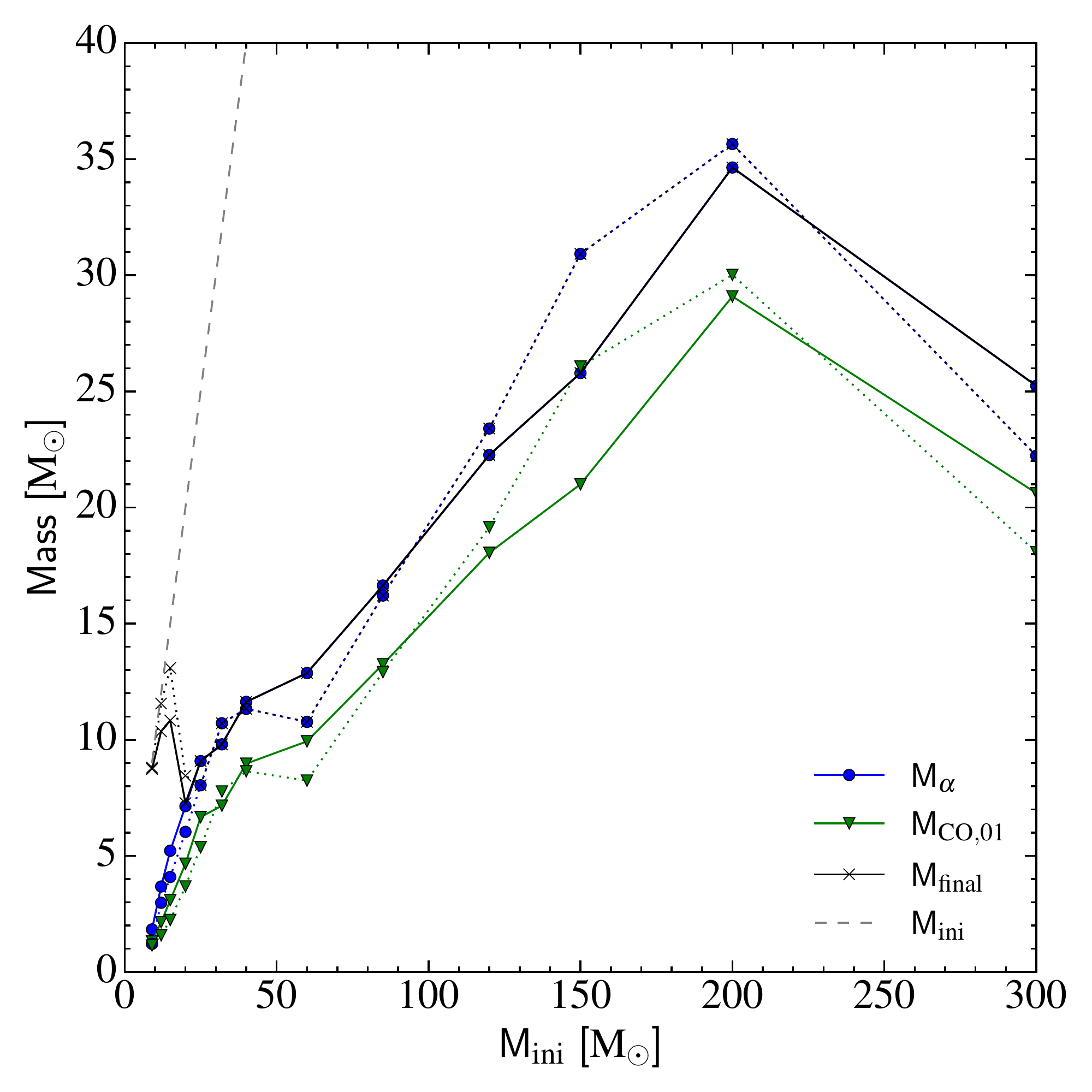}
    \caption{Helium-core ($M_{\alpha,01}$ defined as the mass coordinate where the hydrogen mass fraction drops below 1\%) and carbon-oxygen core masses ($M_{\textrm{CO},01}$ defined as the mass coordinate where the helium mass fraction drops below 1\% and $M_{\textrm{CO},20}$ defined as the mass coordinate where the sum of the mass fractions of carbon and oxygen becomes larger than 20\%) of the models. Straight line is for rotating model and dotted line is for non-rotating model. {\it Left}: zoom in on the mass range from 9 to 40\,\msun. {\it Right}: the full mass range from 9 to 300\,
    \msun.}
    \label{fig:finalcoremasses}
\end{figure*}

The final total mass along with core masses of the models are listed in Table\,\ref{Table:MfinMcores} and plotted in Fig.\,\ref{fig:finalcoremasses}. The strong mass loss experienced by high-metallicity stars leads to final total masses being much lower than initial masses for both non-rotating and rotating stars. The maximum final mass in the entire grid is 36\,\msun\ for the 200\,\msun\ non-rotating model. The maximum final total mass for rotating models is very similar (35\,\msun\ for the rotating 200\,\msun\ model). It is interesting to note that further increasing the initial mass of the model does not lead to an increase in the final mass (The 300\,\msun\ models have final masses smaller than 26\,\msun). This is due to the strong luminosity dependence of mass loss rates. While there are still uncertainties related to mass loss (especially in the cool part of the HRD), it is very unlikely that stars would be able to retain more than 40\,\msun\ at super-solar metallicity and this would also represent an upper limit for black hole masses coming from single stars {at this metallicity}\footnote{Much larger BH masses
are predicted at lower metallicities \citep[see e.\,g.][]{Farrell2021,Vink2021,Umeda2020}.}

. 
Related to this, the models do not predict any pair-instability supernova at super-solar metallicity.

Comparing the total and helium core ($M_\alpha$) masses, one can see that mass loss is strong enough in stars above 20\,\msun\ for rotating stars (25\,\msun\ for non-rotating stars) to remove the entire hydrogen-rich envelope. Our models thus predict SNII below 20\,\msun\ for rotating stars (25\,\msun\ for non-rotating stars) and SNIb above that. We provide two values for the carbon-oxygen core masses. $M_{\textrm{CO},01}$ is defined as the mass coordinate where the helium mass fraction drops below 1\%. It roughly corresponds to the maximum mass reached by the convective core during core helium burning. It is also the location of a steep density gradient at the edge of that core. This gradient will help the SN shock-wave to eject material above that point and it is thus our recommended value for the CO core mass of our models. There are different ways of determining core masses from stellar models \citep[see e.\,g.][]{Hirschi2004}. We thus also provide another measure of the CO-core mass, $M_{\textrm{CO},20}$, defined as the {first} mass coordinate { moving from the surface to the center} where the sum of the mass fractions of carbon and oxygen becomes larger than 20\%. In hydrogen-rich models, this definition falls in between $M_\alpha$ and $M_{\textrm{CO},01}$ . For H-free models, $M_{\textrm{CO},20}$ usually includes the helium burning shell layer, which is composed of helium, carbon and oxygen in various ratios and is thus equal to $M_\alpha$. One could wrongly conclude that there is no helium left in these models. This discussion also demonstrates that it is important to use a comparable definition to compare different grids of models. Comparing $M_\alpha$ and $M_{\textrm{CO},01}$, we see that the models all retain several solar masses of helium-rich material (the helium surface abundance is given in Table\,\ref{TabListModels}). It is still debated \citep[see e.\,g.][and references therein]{2020A&A...642A.106D} whether some (and how much) helium can be hidden in SNIc. If helium cannot be hidden, then our models would not predict any SNIc at super-solar metallicity, only SNII up to about 20\,\msun\ for rotating stars (25\,\msun\ for non-rotating stars) and SNIb above that (using the mass loss prescriptions described in Sect.\,\ref{sec:ingredients}).

\subsection{Comparison to Geneva grids at other metallicities}
\begin{figure*}
    \centering
    \includegraphics[width=.5\textwidth]{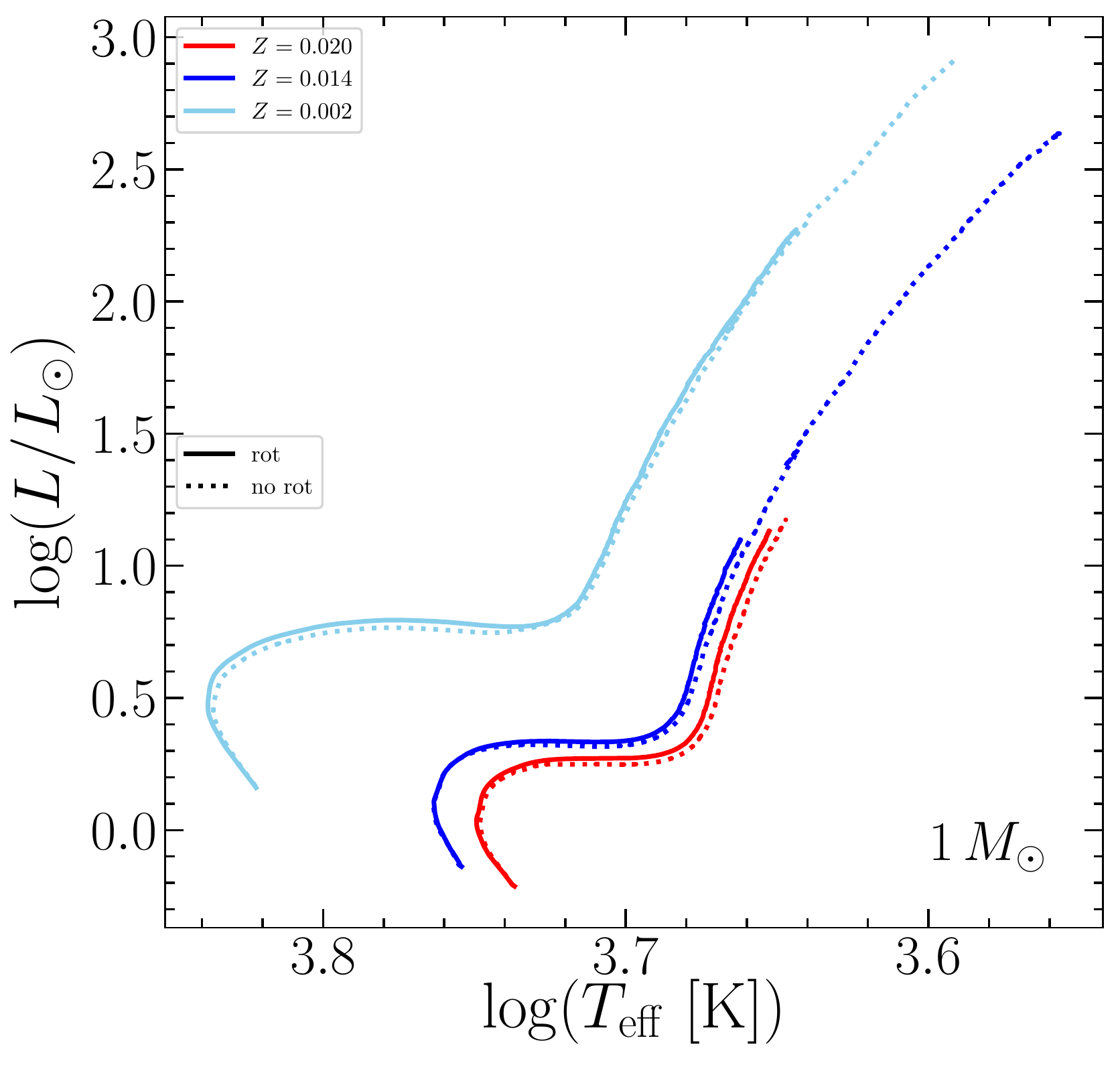}\hfill\includegraphics[width=.5\textwidth]{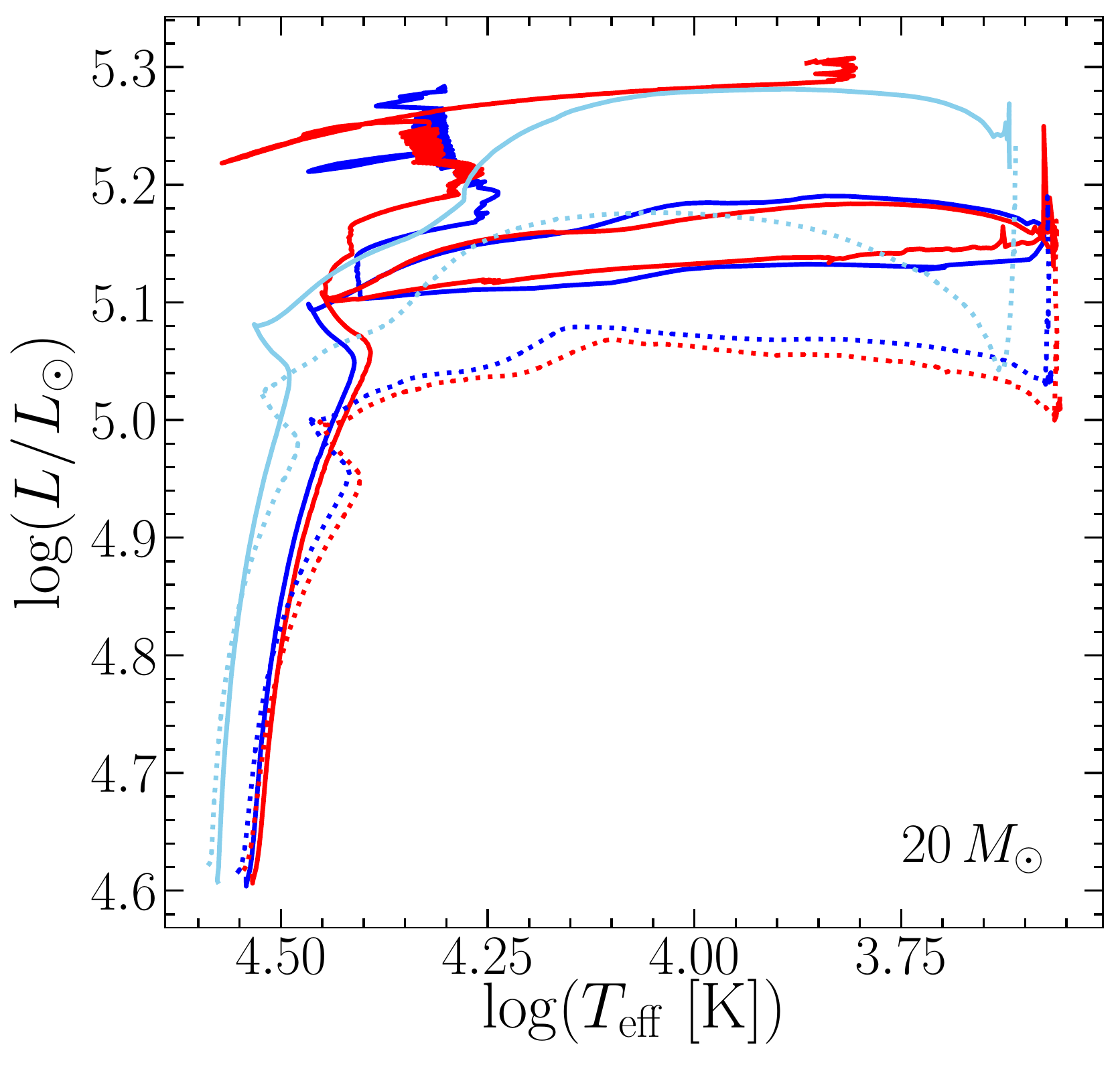}
    \caption{HR diagrams for $1$ ({\it left}), and $20\,\msun$ ({\it right}) models with and without rotation, at $Z=0.020$ (red), $Z=0.014$ \citep[dark blue][]{Ekstrom2012}, and $Z=0.002$ \citep[sky blue][]{Georgy2013b}.}
    \label{fig:HRD_Zcomp}
\end{figure*}

Given the modest difference in metallicity between the super-solar and solar metallicity models (43\%), it is expected that the models at both metallicities have a qualitatively similar evolution, which is indeed the case when comparing most properties of the grids of models. As discussed in Sect.~\ref{sec:intro}, it is nevertheless very useful to have a grid of models tailored to the metallicity of the Galactic Centre to first confirm expectations and second avoid the reliance on extrapolation of model properties to a different metallicity. 
We expect the effects of metallicity in super-solar metallicity models to be in the opposite direction to the effects in low-metallicity models. This is confirmed for the evolutionary tracks in the HRD as can be seen for the 1 and 20\,\msun\ models in Fig.\,\ref{fig:HRD_Zcomp} presenting non-rotating and rotating models at $Z=0.020, 0.014$ and 0.002. The figure shows that the higher the metallicity, the cooler and slightly less luminous the tracks on the MS (explained mainly by the higher opacity at higher metallicity). This leads to slightly longer MS lifetimes (by 20\% or less) for super-solar models compared to solar metallicity models.

\begin{figure}
    \centering
    \includegraphics[width=.45\textwidth]{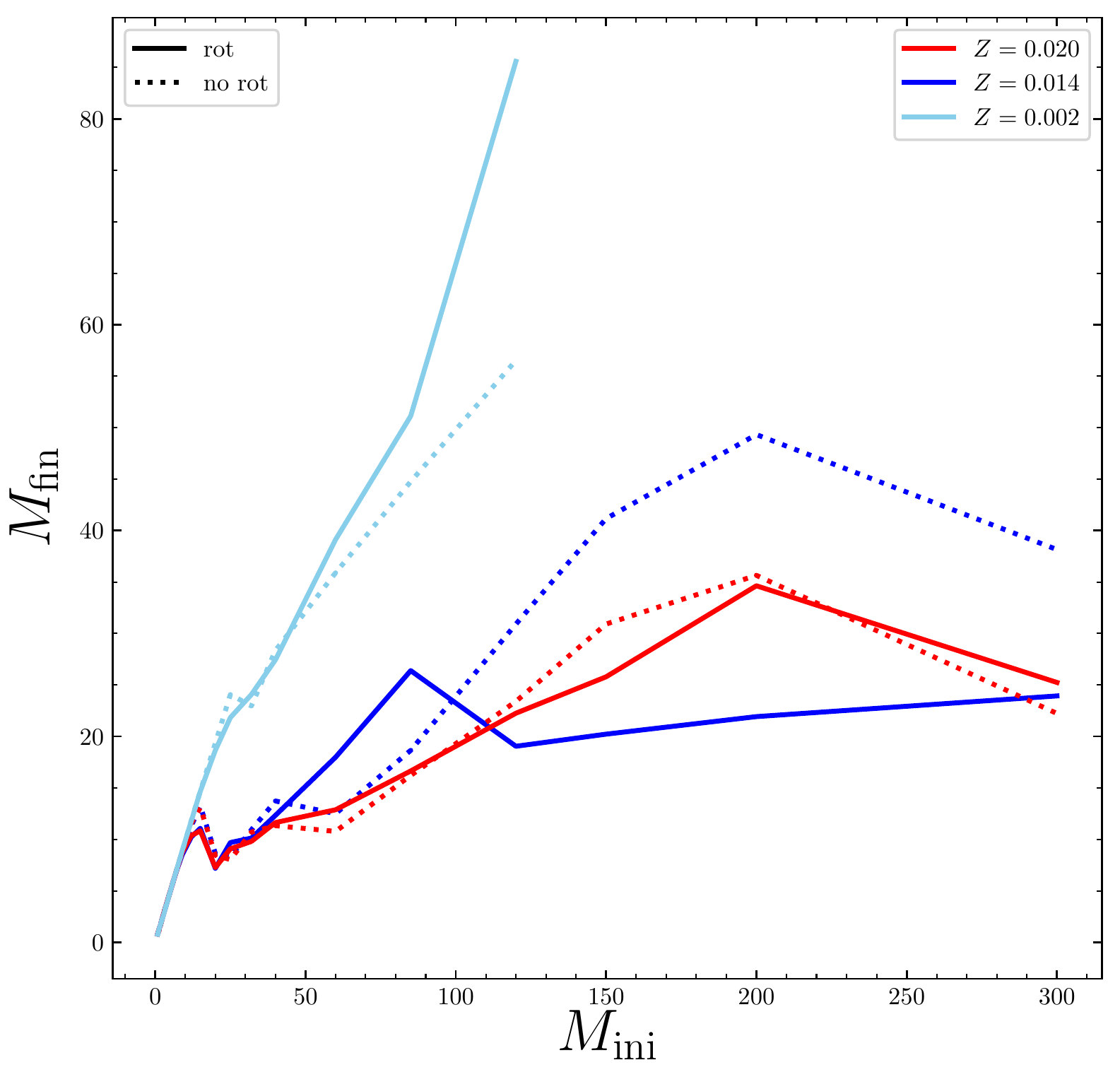}
    \caption{Final mass as a function of initial mass for models with and without rotation at  $Z=0.020$ (red), $Z=0.014$ \citep[dark blue][ for $M_\text{ini}>120\,$\msun]{Ekstrom2012,Yusof2013}, and $Z=0.002$ \citep[sky blue][]{Georgy2013b}.}
    \label{fig:MfinMiniZ}
\end{figure}

Mass loss is one of the properties of stellar models most affected by metallicity. While the dependence of mass loss on metallicity varies according to the location of the star in the HRD (see Sect.\,2 for details), the general trend is that mass loss is higher at higher metallicity, which leads to lower final masses. The final masses of the $Z=0.020$ models are compared to models at $Z=0.014$ and 0.002 in Fig.\,\ref{fig:MfinMiniZ}. While super-solar metallicity models lose much more mass than low-$Z$ models, final masses are similar to solar metallicity models up to 30\,\msun. This can be explained by several factors: first the proximity in metallicity between the two grids, second the fact that the models do not include a metallicity dependence for $\log_{10} (\teff / K) < 3.7$, and third the dependence of mass loss rates on luminosity. The second factor plays an important role for stars in the 15 to 30\,\msun\ range since stars in this range lose most mass as very cool RSGs. The third factor plays a dominant role for very massive stars as discussed above. Indeed, very massive stars are so luminous that they lose a lot of mass. This reduces the luminosity of the star, which in turns reduces its mass loss. The maximum final mass of the super-solar models around 35\,\msun\ is lower than the maximum mass of 49.3\,\msun\ \citep[for the 200\,\msun\ from][]{Yusof2013} retained by non-rotating solar metallicity models while it is higher than the maximum mass reached by rotating solar metallicity models of 26.4\,\msun\ \citep[for the 85\,\msun\ model from][]{Ekstrom2012}. The maximum mass retained depends both on the evolutionary path taken by the VMS models (non-rotating models reach cooler temperatures than rotating ones) and the mass loss experienced during the various phases. An important finding from our models is that starting from an even higher initial mass would not allow high-metallicity stars to produce more massive black holes (no black hole masses predicted above 50\,\msun\ for solar or higher metallicities). This confirms that at high metallicity mass loss is the major process determining the maximum mass of black holes from single stars. The models would thus not predict pair-instability SNe at solar or super-solar metallicities. As can be seen from the $Z=0.002$ models, this is not the case at sub-solar metallicities \citep[see also][and references therein]{Eggenberger2021,Higgins2021}.

\begin{figure*}
    \centering
    \includegraphics[width=.5\textwidth]{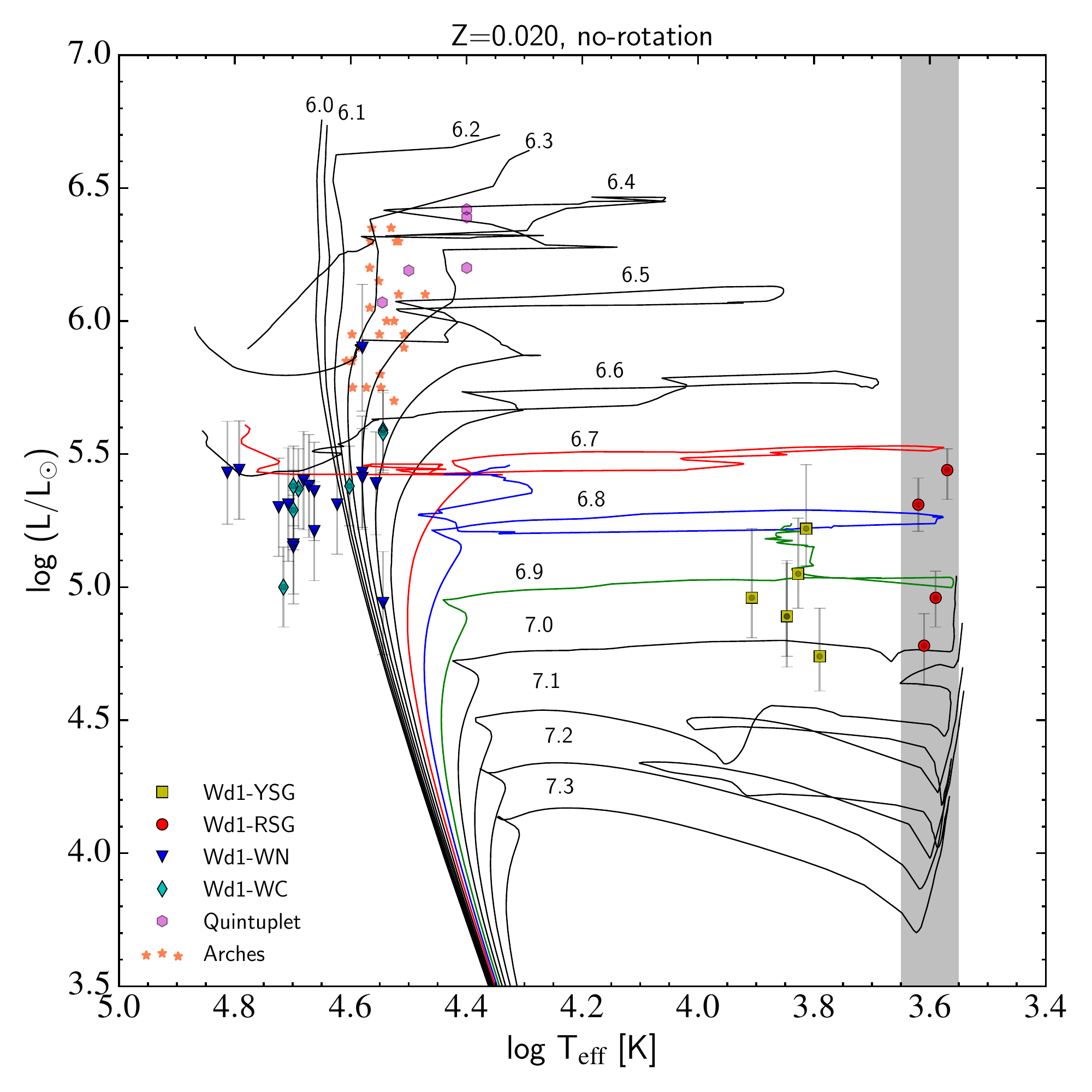}\includegraphics[width=.5\textwidth]{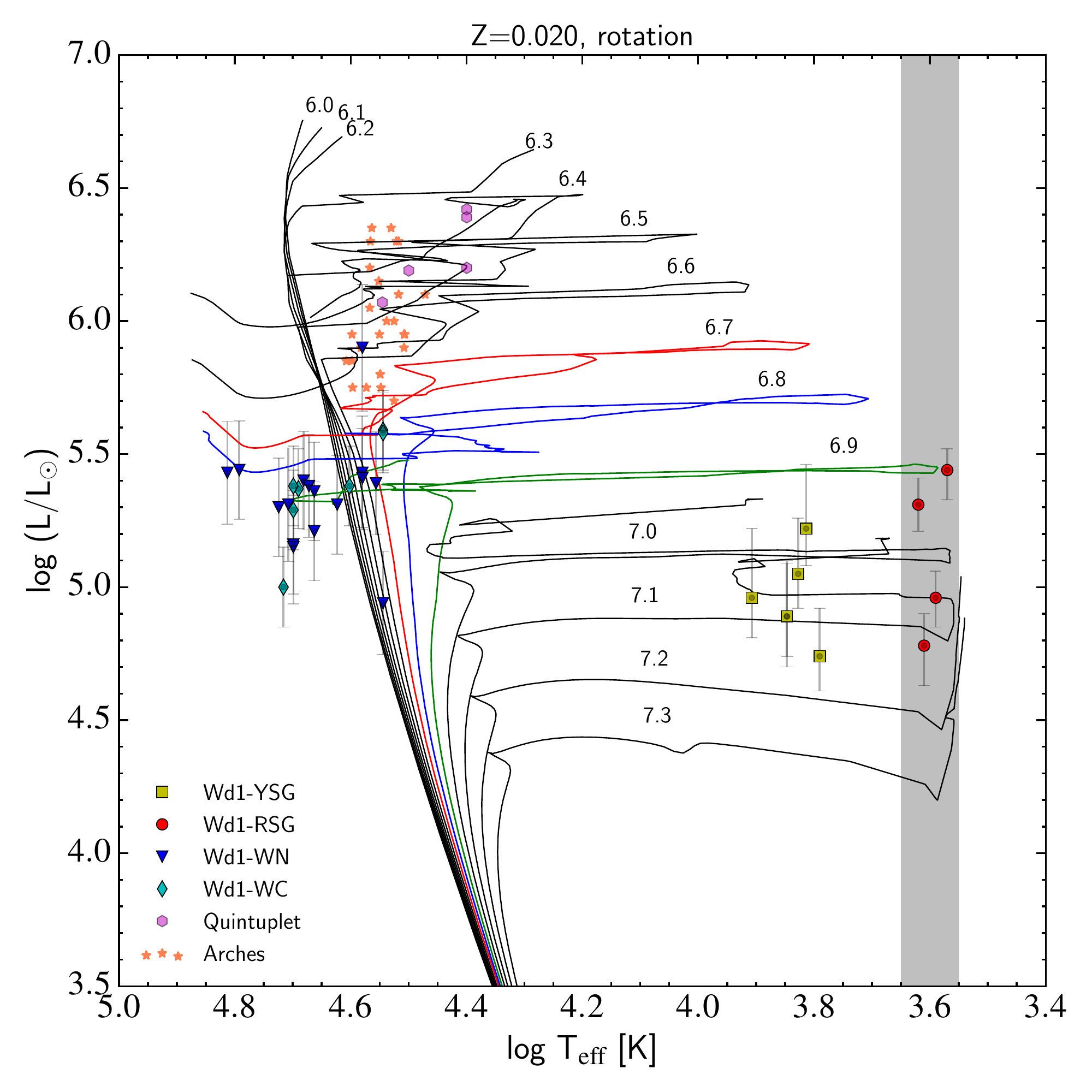}
    \caption{HR diagram comparing cool supergiants in Wd1 \citep[red circles for RSG, yellow squares for YSG;][]{beasor2021age} and hot, luminous stars in the Wd 1 \citep[blue inverted triangles for WN stars, cyan diamonds for WC stars;][]{Rosslowe2016}, Arches \citep[orange stars;][]{Martins2008} and Quintuplet \citep[pink hexagons;][]{Liermann2010} clusters to the supersolar metallicity isochrones for non-rotating ({\it left}) and rotating ({\it right}) models. The number along or at the end of the isochrones corresponds to the log$_{10}$ of the age of the isochrone. Isochrones with log$_{10}(\text{age/yr}) = $ 6.7, 6.8 and 6.9 are coloured in red, blue and green, respectively to facilitate the comparisons. The grey shaded area indicates the temperature range for the observed RSG stars.}
    \label{fig:isochrones}
\end{figure*}
\section{Comparison to observations}\label{sec:obs}
As discussed in Sect.~\ref{sec:intro}, there are several massive young clusters in the inner Galactic disk. The best studied massive young star cluster in the inner Galactic disk 
is Westerlund~1 (Wd1), at a distance of $\sim$4 kpc 
\citep{beasor2021age}, while there are also several older massive 
clusters at the end of the Galactic Bar which are rich in red 
supergiants \citep{Davies2009}. Wd1 \citep{Westerlund1961} is perhaps 
the richest young star cluster within the disk of the Milky Way. 
\citet{clark2005massive} first highlighted its exceptional population of both hot and cool evolved massive stars, arising from its high cluster 
mass, $\sim 10^{5} M_{\odot}$. The simultaneous presence of Wolf-Rayet 
stars (WN and WC) and cool supergiants led to a preferred age of 
$\sim4-5$ Myr \citep{crowther2006} based on predictions from single star evolutionary models, although \cite{beasor2021age} have recently 
reassessed the age of Wd1 on the basis of its cool supergiant population and argue for an older age of $\sim$10\,Myr. Unfortunately, the 
metallicity of Wd1 is not known, since gas associated with the formation of the cluster has been dispersed. This prevents standard nebular 
diagnostics, and the usual present day stellar diagnostics (iron lines 
in blue spectra of B-type stars) are inaccessible owing to high 
foreground extinction \citep [$A_{\rm V}\sim$ 13 
mag,][]{clark2005massive}.

Within the Galactic Centre \citep{GalCentre2019}, there are several young high mass ($\geq 10^{4} M_{\odot}$) clusters including the 
Arches, Quintuplet and Galactic Centre clusters, plus a rich 
massive star population within the Central Molecular Zone 
\citep{Clark2021}.

The Arches cluster is the youngest, densest star cluster in the 
vicinity of the Galactic Centre. It was discovered independently by \cite{nagata1995AJ} and \cite{cotera1996ApJ}.
It hosts a rich population of O stars and hydrogen rich WN stars 
\citep{Martins2008, clark2019a&a}, such that its age is 2--3 Myr. 
The  Quintuplet cluster is somewhat older than the Arches since it 
hosts late O 
supergiants, WC and WN-type Wolf-Rayet stars plus Luminous Blue 
Variables 
(LBVs), with an age of 3--5 Myr from comparison with single star 
models \citep{Liermann2009, Clark2018}. Standard nebular and 
stellar  abundance  diagnostics are also challenging for massive 
stars in the Galactic  Centre due to extreme visual extinction, 
although  \citet{Cunha2007} have analysed 
intermediate to high mass cool supergiants in the central cluster 
and Quintuplet to reveal iron abundances 0.10 to 0.15 dex higher 
than the solar value, with [O/Fe]$\sim$ 0.2 dex. 
\citet{Najarro2009, Najarro2014} have 
obtained similar results for selected early-type stars in the 
Quintuplet cluster.

We will mainly compare our super-solar metallicity models to the massive stars observed in Wd1 since it is the best studied young metal-rich cluster. We will also briefly compare our models to observed stars in the Arches and Quintuplet clusters.

In order to estimate the ages of these clusters, we compare the isochrones of our super-solar models to the observed massive star populations in Fig.\,\ref{fig:isochrones}. Starting with the Arches and Quintuplets clusters, we see that luminous stars in these clusters \citep{Liermann2010, Martins2008} fall between the isochrones with log$_{10}(\text{age/yr}) = $ 6.3 and 6.5. These values match previous age estimates for the Arches cluster \citep[2-3\,Myr][]{Martins2008, Clark2018A&A} and is close to prior age estimates for the Quintuplet cluster \citep[3-5\,Myr][]{Liermann2012, Clark2018}. Comparing the observations to the evolutionary tracks of our models (not shown here), late WN stars in the Quintuplet cluster have initial masses above 80\,\msun, although neither O stars nor (dusty) WC stars have been subject to quantitative investigation to date, hindering a more refined age determination. For the Arches cluster (and indirectly other Galactic Centre clusters), \citet{Clark2018A&A} have emphasised the sensitivity of  stellar luminosities to the adopted extinction law. The discovery of a very high mass binary system (F2) in the Arches cluster \citep{Lohr2018} favours its youth with respect to alternative interpretations involving the most massive stars being the products of binary evolution \citep{Schneider2014}. 
More detailed studies would be needed to provide precise 
information on these clusters, consequently we will focus on Wd1 
for the rest of the comparisons. 

Extensive spectroscopic studies of the massive star population in 
Wd1 have been undertaken since this cluster first came to 
prominence \citep{clark2005massive}, although in common with the 
Galactic Centre clusters, quantitative spectroscopic results have 
not been undertaken for OB stars in Wd1. The simultaneous presence 
of WR stars and cool supergiants led \citet{clark2005massive} to 
conclude that its age was 3.5--5 Myr. \citet{crowther2006}  
provided estimates of the physical properties of WR stars in Wd1, 
from which a cluster age of $\sim$5 Myr was favoured from 
comparison with predictions from single stars. \citet{Rosslowe2016} undertook a more detailed analysis of the WR population of Wd1, taking into account contributions from binary companions and hot dust emission. \citet{Negueruela2010} favoured log$_{10}(\text{age/yr})$ = 6.7 to 6.8 from a comparison between its rich OB supergiant population and previous generations of Geneva single star models, with its youth reinforced from the detection of high mass eclipsing binaries \citep{Ritchie2010}. In contrast, \citet{beasor2021age} have reassessed the luminosities of cool supergiants in Wd1 incorporating mid-IR photometry to infer
a substantially older age of $\sim$10 Myr. 

From the comparison between the physical properties of WR stars and cool supergiants in Wd1 to new isochrones in Fig.\,\ref{fig:isochrones}, we see that non-rotating models at log$_{10}(\text{age/yr})$ around 6.7 reach the position of the WR stars, although the luminosity of the models is slightly too high. Non-rotating models at log$_{10}(\text{age/yr})$ around 6.9 overlap with the position of the red and yellow supergiants (RSG \& YSG) from \cite{beasor2021age}. Rotating models on the other end, reach the position of the WR stars at log$_{10}(\text{age/yr})$ around 6.8-6.9 while they cover the region occupied by the RSGs and YSGs for log$_{10}(\text{age/yr})$ around 7.0. 
\begin{figure*}
    \centering
    \includegraphics[width=.5\textwidth]{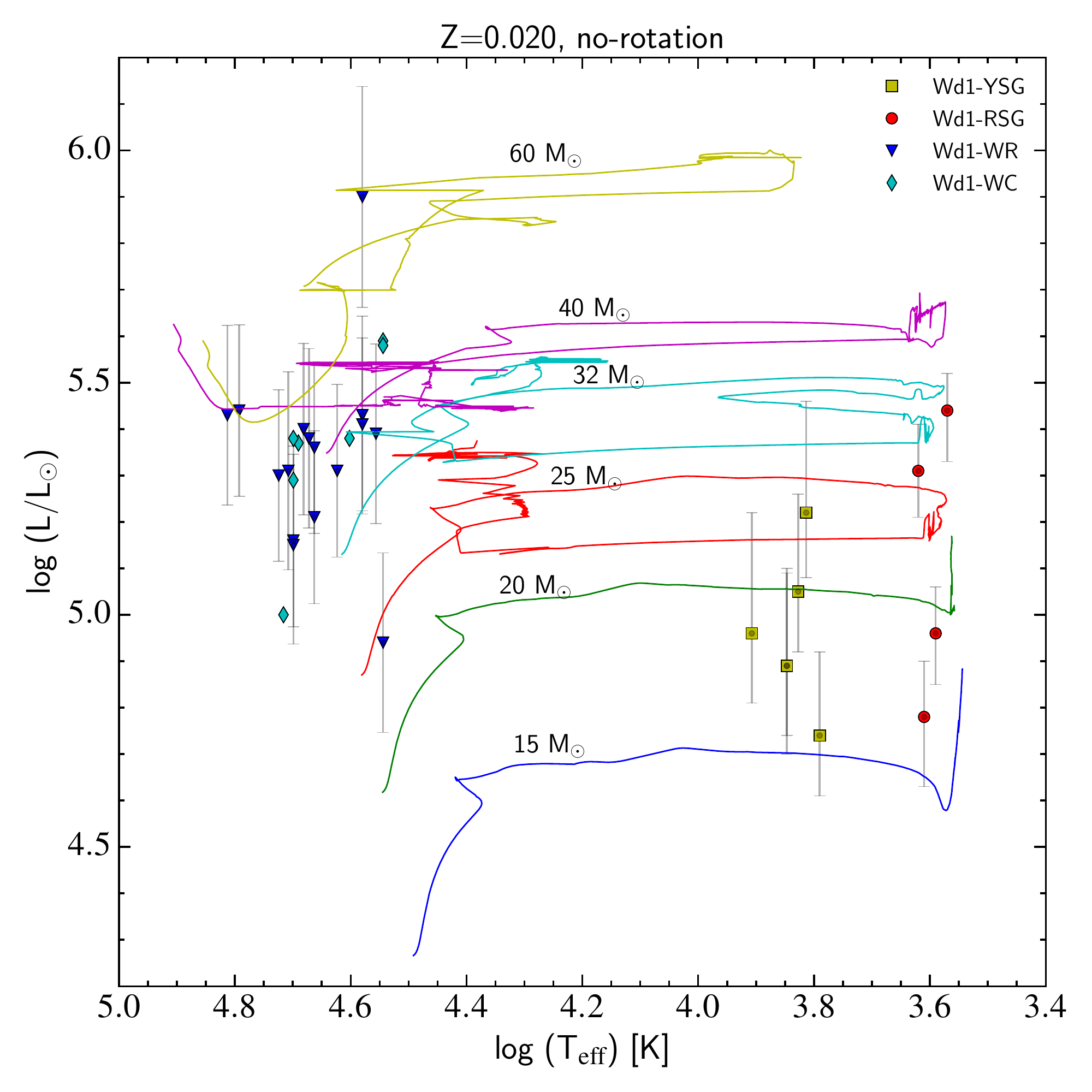}\includegraphics[width=.5\textwidth]{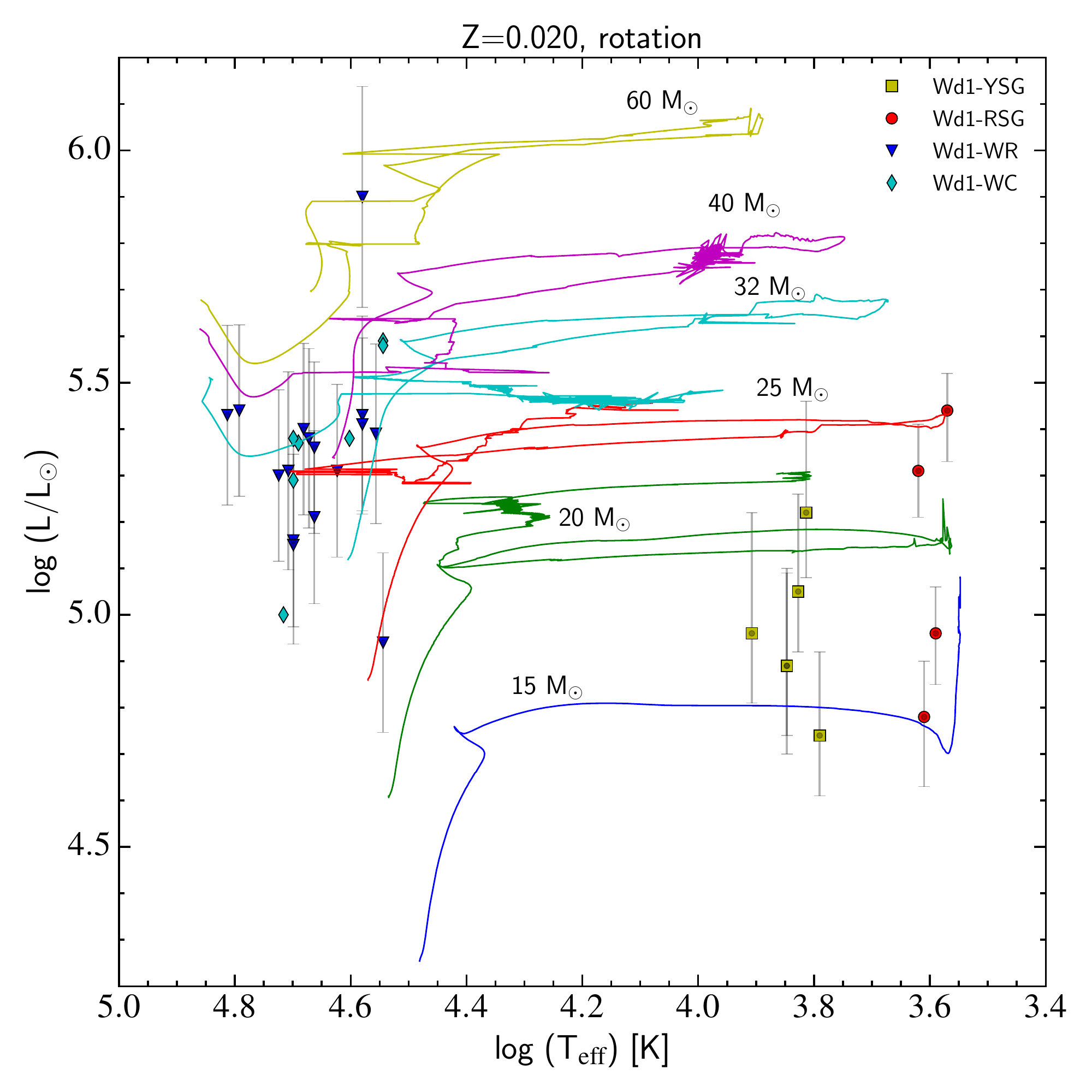}
    \caption{Comparison in the HRD of the evolutionary tracks of the 15 to 60\,\msun, non-rotating ({\it left}) and rotating models ({\it right}) to the Wd1 data (same Wd1 data as in Fig.\,\ref{fig:isochrones}).}
    \label{fig:tracksobs_Westerlund}
\end{figure*}

\begin{figure}
    \centering
    \includegraphics[width=.45\textwidth]{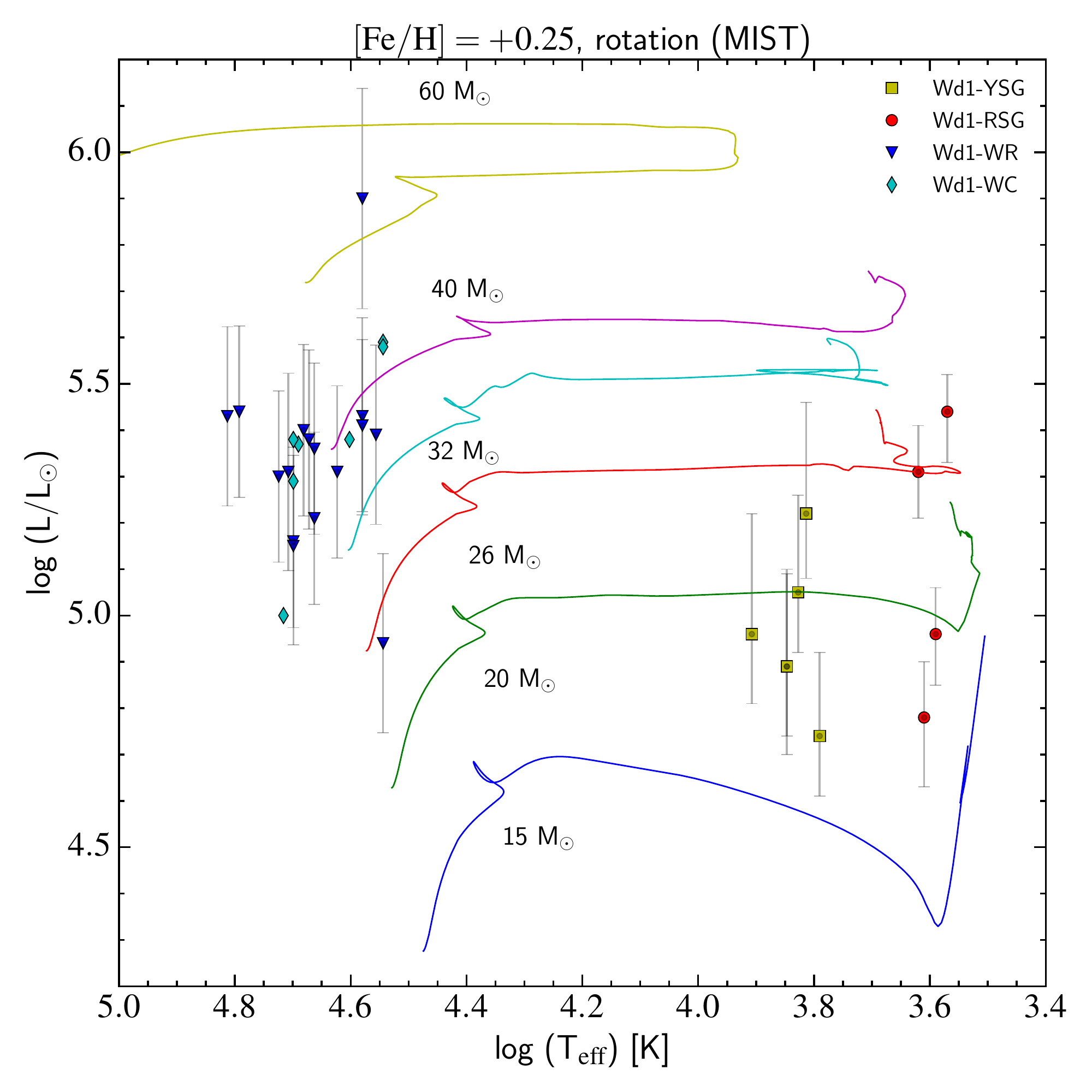}
    \caption{{Comparison in the HRD of MIST evolutionary tracks for the $15-60\,\msun$ 
    rotating models at [Fe/H]$+0.25$ ($Z=0.0254$) to the Wd1 data (same Wd1 data as in Fig.\,\ref{fig:isochrones}).} 
    }
    \label{fig:tracksobs_Westerlund_MIST}
\end{figure}

To find out the initial masses of the models reaching the observed position of Wd1 stars, we compare observations to evolutionary tracks of our super-solar models between 15 and 60\,\msun\ in Fig.\,\ref{fig:tracksobs_Westerlund}. We see that non-rotating models with $M_\text{ini}$ above 25\,\msun\ reach the position of the Wolf-Rayet (WR) stars in Wd1 (though again the luminosity of the models is slightly higher than that of the observed WR stars), while non-rotating models with $M_\text{ini}$ between 15 and 32\,\msun \ overlap with the position of the red and yellow supergiants (RSG \& YSG). Rotating models with $M_\text{ini}$ above 20\,\msun reach the position of the Wolf-Rayet (WR) stars in Wd1 while they cover the region occupied by the RSGs and YSGs for $M_\text{ini}$ between 15 and 25\,\msun. 

As already discussed in Sect.\,\ref{sec:SSmodels}, the mass range of these models corresponds to the transition between stars ending as RSGs (and SNII) and those ending as WRs (SNIb/c). In both the models and Wd1 stars, this transition occurs for \llsun\ = $5-5.5$ (possibly at a slightly lower luminosity in Wd1 stars compared to the models). 

\begin{figure*}
    \centering
    \includegraphics[width=.5\textwidth]{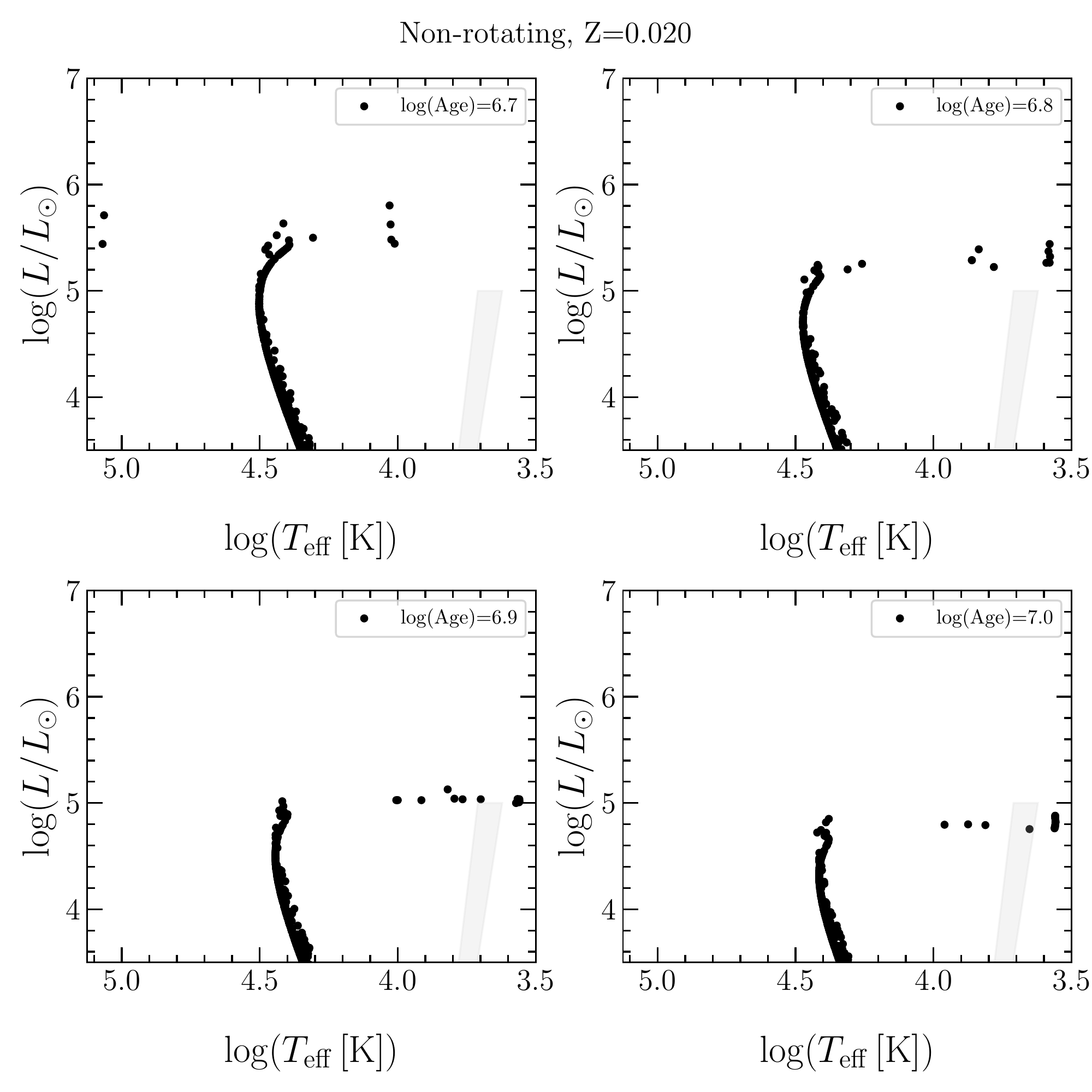}\includegraphics[width=.5\textwidth]{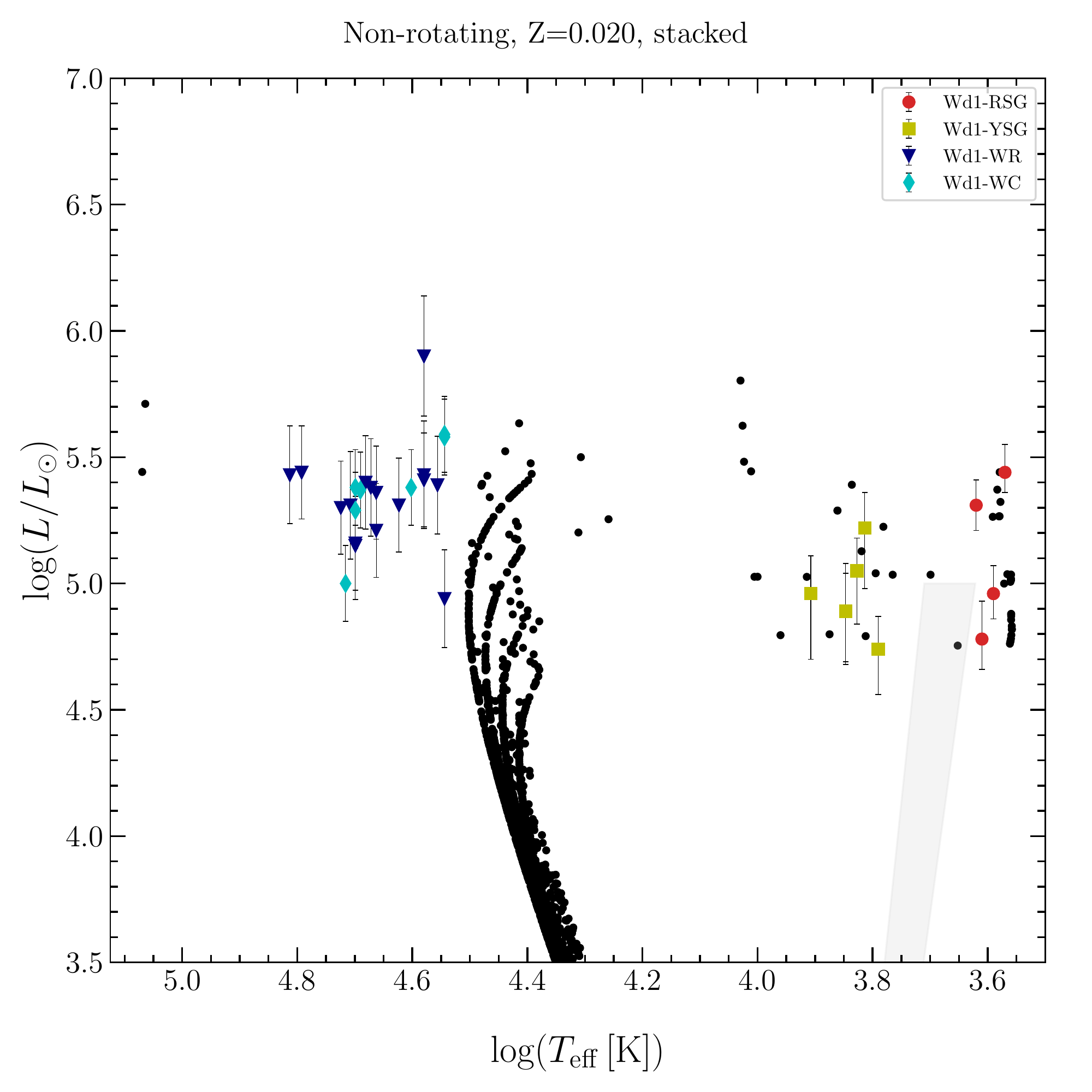}
    \caption{HRD of synthetic clusters with log$_{10}(\text{age/yr})=$ 6.7, 6.8, 6.9, 7.0, at $Z=0.020$, each with 10,000 initial stars, without rotation. {\it Left:} Individual 10,000-star clusters considering instantaneous star formation. {\it Right:}  Cluster of 40,000 initial stars combining (stacking) the stars from the four 10,000-star individual clusters on the left, corresponding to a star formation episode lasting 5\,Myr.}
    \label{fig:Wd1_clusters_norot}
\end{figure*}

\begin{figure*}
    \centering
    \includegraphics[width=.5\textwidth]{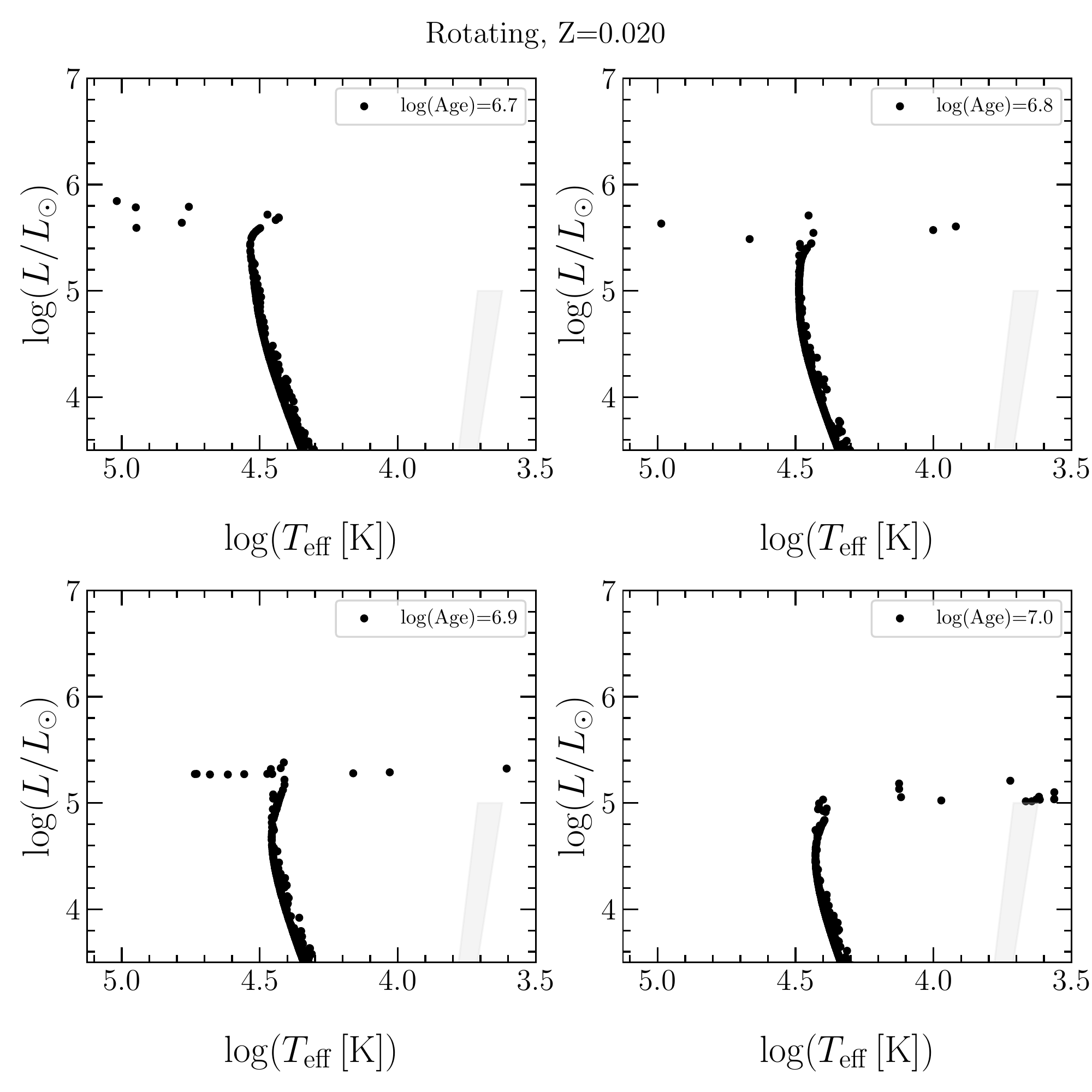}\includegraphics[width=.5\textwidth]{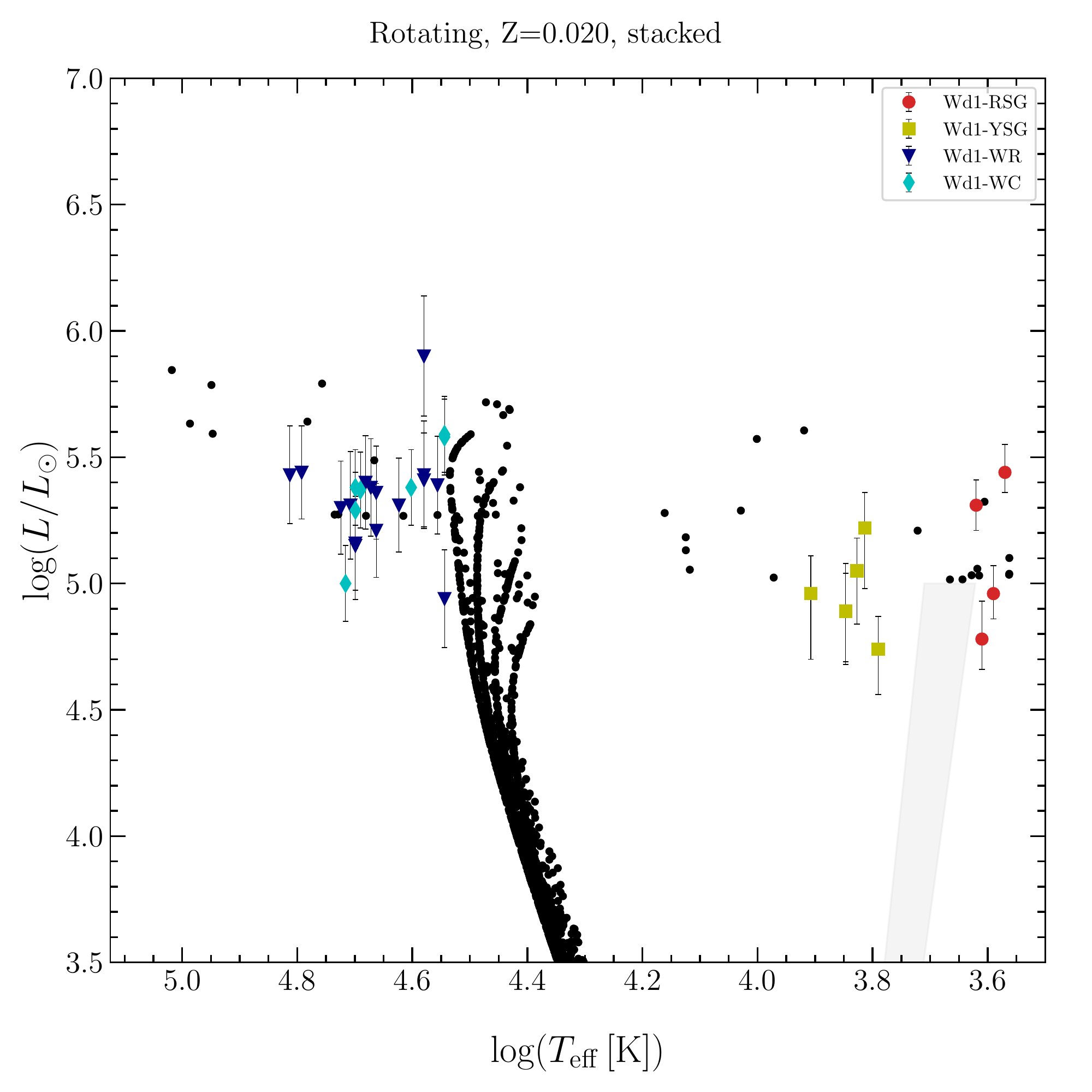}
    \caption{Same as Fig.\,\ref{fig:Wd1_clusters_norot} but for rotating models.}
    \label{fig:Wd1_clusters_rot}
\end{figure*}

\begin{figure*}
    \centering
    \includegraphics[width=.5\textwidth]{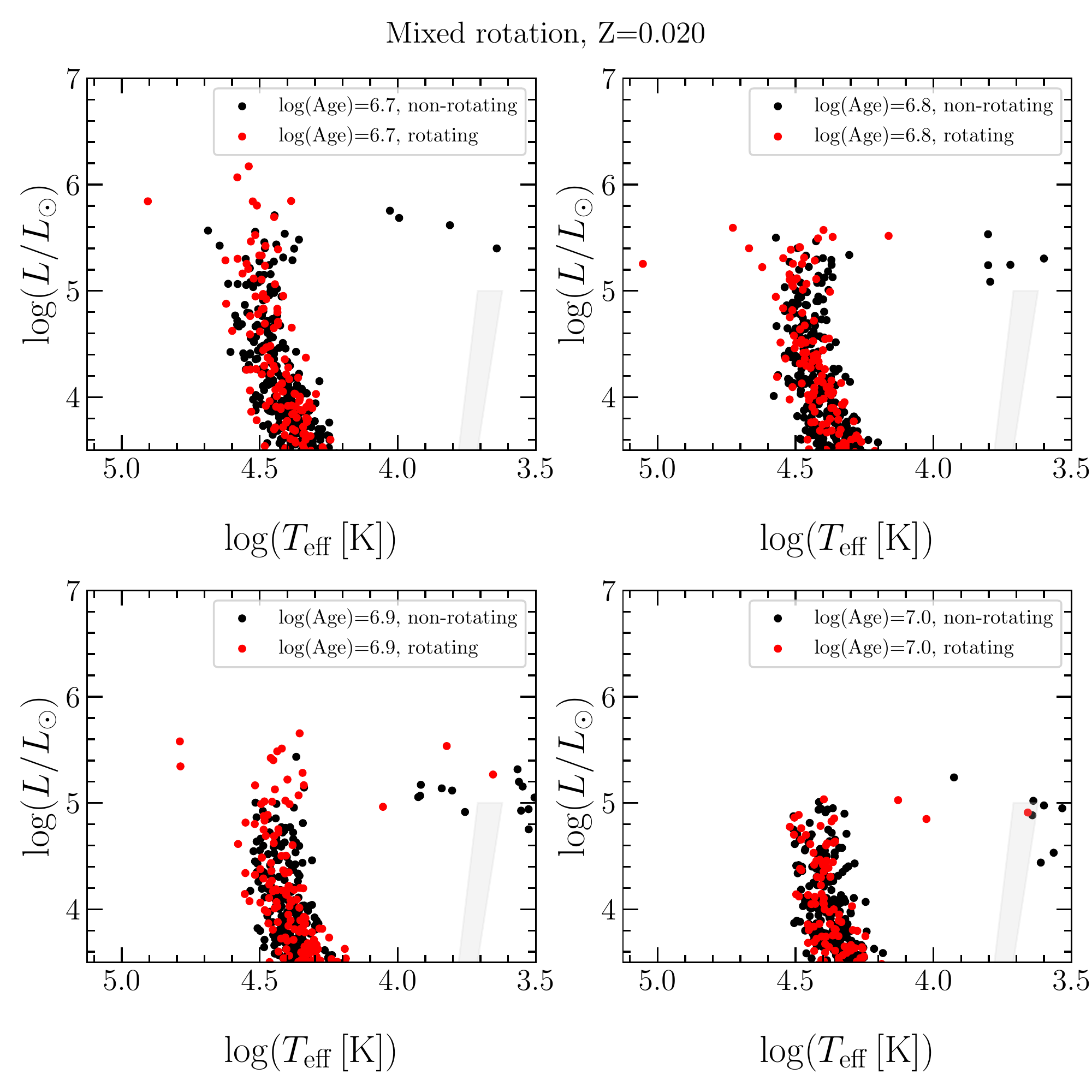}\includegraphics[width=.5\textwidth]{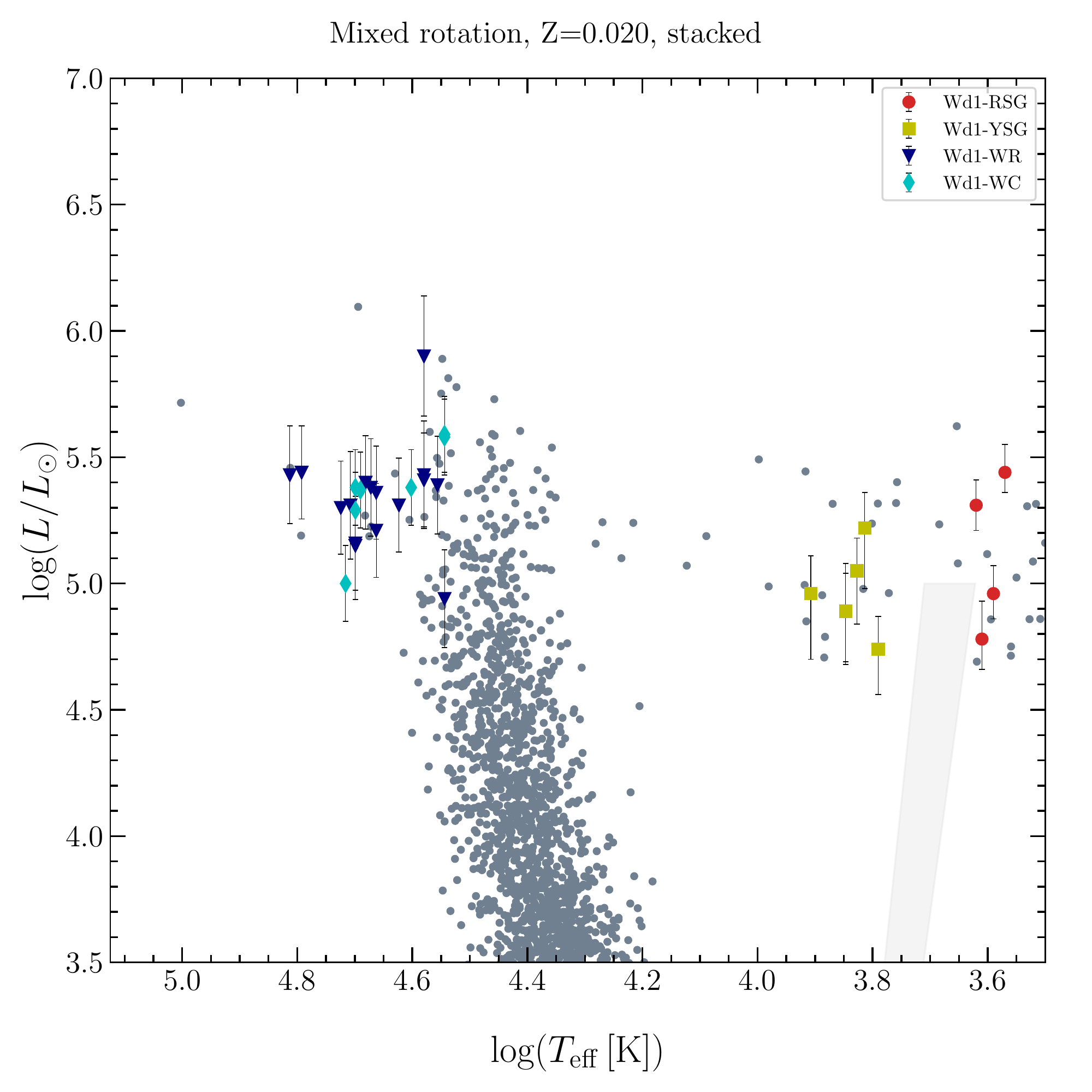}
    \caption{HRD of synthetic clusters of $\log{\rm (age/yr)}$ 6.7, 6.8, 6.9, 7.0, at Z=0.020, each containing 10000 initial stars with mixed rotation: 7000 without rotation (black dots) and 3000 with rotation (red dots). We added gaussian noise on $L$ and $T_{\rm eff}$ to simulate observed stars ($\sigma_L=0.2~$dex, $\sigma_{T_{\rm eff}}=0.05~$dex). {\it Left:} Individual 10,000-star clusters considering instantaneous (or very brief) star formation. {\it Right:}  Cluster of 40,000 initial stars combining the stars from the four 10,000-star individual clusters on the left, corresponding to a star formation episode lasting 5\,Myr.}
    \label{fig:Wd1_clusters_mix}
\end{figure*}

To take the comparison one step further, we used the SYCLIST tool \citep[see][for details]{Georgy2014} to generate synthetic clusters out of the super-solar models.
The estimated total stellar mass of Wd1 is $\sim 10^5M_\odot$. Using a Salpeter IMF with the lower and upper mass bounds from the grids (0.8 and 300\,\msun, resp.) yields an average mass of $\sim 2.6M_\odot$. We thus generated clusters initially containing 40,000 stars in total. To take into account the age estimates ranging from about 5 to 10\,Myr from the above comparison to isochrones \citep[as well as age determinations from the literature ][]{clark2005massive, crowther2006, beasor2021age} and the possibility of a cluster formation event lasting a few million years, we computed four clusters with log$_{10}(\text{age/yr})=6.7, 6.8, 6.9, 7.0$, each with 10,000 initial stars. We construct such clusters for both the non-rotating (Fig.\,\ref{fig:Wd1_clusters_norot}) and the rotating (Fig.\,\ref{fig:Wd1_clusters_rot}) models. Note that while there are initially 40,000 stars in the clusters, the most massive stars will have died by the age at which we compute the clusters. Thus the total mass of the synthetic clusters is slightly below $10^5M_\odot$. We also make a more realistic cluster with mixed rotation, initially containing $7000\times4$ non-rotating and $3000\times4$ rotating stars (Fig.\,\ref{fig:Wd1_clusters_mix}). For this `mixed rotation' cluster we add Gaussian noise on $L$ and $T_{\rm eff}$ to simulate observed stars ($\sigma_L=0.2~$dex, $\sigma_{T_{\rm eff}}=0.05~$dex).

Figure\ \ref{fig:Wd1_clusters_norot} shows that the non-rotating clusters are able to broadly reproduce the observed RSGs and YSGs in Wd1. However they produce too few Wolf-Rayet stars and those produced possess $T_{\rm eff}$ and \llsun\ about 0.3 dex too high compared to observations (most visible in the {\it left} panels). On the other hand, the rotating clusters (Fig.\,\ref{fig:Wd1_clusters_rot}) yield a better agreement with the observed Wolf-Rayet population. This is explained by the rotating 20-25\,\msun\ models becoming WR stars with \llsun \ $\sim 5.2$ (especially around log$_{10}(\text{age/yr})=6.9$). Rotating models predict fewer RSG/YSG than non-rotating models and the predicted ``YSG'' have higher effective temperatures than typical
YSGs (note however that the effective temperature of the YSG/RSGs in Wd1 is not very precise so one cannot draw firm conclusions on this point).
Finally, the cluster with mixed rotation (right panel of Fig. \ref{fig:Wd1_clusters_mix}) is able to qualitatively reproduce the existence of various evolved sub-types at the observed luminosities. 

The comparison is not perfect, however, especially when considering the relative number of stars in the various sub-types. Indeed, the synthetic cluster contains more RSG and YSG and less WR stars than what is observed. A perfect match was not expected for several reasons. First, the parameters chosen for the synthetic clusters are based on estimates for the total cluster mass, the star formation rate and the distribution of initial stellar rotation and these are uncertain. For example, the cluster formation history of Westerlund-1 is probably not as simple as four independent and identical star formation episodes. If, for instance, its star formation rate (SFR) increased over time, we should expect to see relatively more WR stars and fewer RSG and YSG stars compared to our constant SFR cluster. 
Second, the grid of models only includes single stars so we would expect binary interactions to contribute to the WR stars. This being said, \citet{beasor2021age} are unable to fit their observations with a single-aged BPASS model \citep{BPASS2018}.
Third, another uncertainty concerns mass loss. In particular, mass loss prescriptions on the cool side of the HRD are all empirical (see Sect\,\ref{sec:ingredients} for details).
\citet{beasor2021age} compare their observations to the MIST isochrones of \citet{Choi2016}. The MIST and Geneva isochrones are very different above \llsun $\gtrsim 5$. 
The MIST isochrones never reach effective temperatures above log$_{10} (T_\text{eff})>4.2$ for \llsun \ $< 6.0$ (or ages larger than 4.5 Myr), whereas our isochrones cross back the MS at \llsun \ $\sim 5.5$ (or below), for our non-rotating models. 
Evolutionary tracks of MIST models for the $15-60\,\msun$ 
rotating models at [Fe/H]$+0.25$ ($Z=0.0254$) are shown in Fig.\,\ref{fig:tracksobs_Westerlund_MIST}. These show that these MIST super-solar metallicity models up to 40\,$\msun$ never leave the RSG phase (even considering rotation) so retain at least part of their H-rich envelope.  
The MIST stellar models have various ingredients that are different from the Geneva models used here (e.g. different implementation of rotation-induced mixing) so it is not straightforward to extract the impact of the mass loss uncertainties. 
This being said, the difference between MIST and GENEVA models in the $20-40\,\msun$ mass range is most likely dominated by differences in mass loss in the RSG phase.
The comparison to Wd1 provides support for significant mass loss in the RSG phase for this mass range, thus probably for an enhanced mass loss rate when the sub-surface or surface layers approach the Eddington limit, which is implemented in our models (see Sect.\,\ref{sec:ingredients} for details). Generally, this confirms the importance of mass loss and of the related uncertainties for evolution of massive stars. 
Taking all these factors into considerations, the present grid of 
single star models (with its physical ingredients described in 
Sect.\,\ref{sec:ingredients}) is able to reproduce the Wd1 evolved 
populations rather well, at least qualitatively.

\section{Conclusions}\label{sec:conclusions}
In this paper, we present a grid of stellar models at super-solar metallicity ($Z=0.020$) covering a wide range of initial masses from 0.8 to 300\,\msun. This grid extends the previous grids of Geneva models at solar and sub-solar metallicities \citep{Ekstrom2012,Eggenberger2021,Georgy2013b,Groh2019a,Murphy2021} and thus uses the same physical ingredients and metallicity dependencies. A metallicity of $Z=0.020$ was chosen to match that of the inner Galactic disk. After presenting the models, we compare them to Geneva grids at other metallicities and several massive young stellar clusters near the Galactic centre, Westerlund 1 (Wd1), in particular. 

A modest increase of 43\% (=0.02/0.014) in metallicity compared to solar models means that the models evolve similarly to solar models but with slightly larger mass loss rates. Mass loss limits the final total mass of the super-solar models to 35\,\msun, even for stars with initial masses much larger than 100\,\msun. Thus the models would predict neither pair-instability supernovae nor BHs above 35\,\msun\ at super-solar metallicity. Furthermore, mass loss is strong enough in stars above 20\,\msun\, for rotating stars (25\,\msun\, for non-rotating stars) to remove the entire hydrogen-rich envelope. Our models thus predict SNII below 20\,\msun\, for rotating stars (25\,\msun\, for non-rotating stars) and SNIb (possibly SNIc) above that.

We computed both isochrones and synthetic clusters to compare our super-solar models to the Wd1 massive young cluster.  
A synthetic cluster combining rotating and non-rotating models with an age spread between $\log{\rm (age/yr)}=$ 6.7 and 7.0 is able to reproduce qualitatively the observed populations of WR, YSG and RSG stars in Wd1. In particular, the models are able to reproduce the simultaneous presence of WR, YSG and RSG stars at \llsun\ 5-5.5. The quantitative agreement is not perfect though and we discuss the likely causes: synthetic cluster parameters, binary interactions and mass loss and the related uncertainties. In particular, mass loss in the cool part of the HRD plays a key role 
\citep[as demonstrated by the different predictions between this study and][]{Choi2016}. Furthermore, larger convective boundary mixing supported by various studies \citep[see e.\,g.][and references therein]{Scott2021} would likely lower the minimum initial mass of a single star to produce a WR star.

\section*{Acknowledgements} N Yusof and HA Kassim acknowledge the Fundamental Research Grant Scheme grant  number FP042-2018A and FP045-2021 under Ministry of Higher Education Malaysia. HAK would like to thank Astrophysics Group, Keele University for hosting his sabbatical where part of this works has been done. RH acknowledges
support from the World Premier International Research Centre Initiative (WPI
Initiative, MEXT, Japan), STFC UK, the European Union’s Horizon 2020 research and innovation programme under grant agreement No 101008324 (ChETEC-INFRA) and the IReNA AccelNet Network of Networks, supported by the National Science Foundation under Grant No. OISE-1927130. This article is based upon work from the ChETEC COST Action (CA16117), supported by COST (European Cooperation in Science and Technology). PE, SE, CG, YS and GM have received funding from the European Research Council (ERC) under the European Union's Horizon 2020 research and innovation programme (grant agreement No 833925, project STAREX). JHG, EF and LM wish to acknowledge the Irish Research Council for funding this research.

\section*{Data availability} An interactive tool to access the models can be found at this address: \url{https://www.unige.ch/sciences/astro/evolution/en/database/}. Additional data requests can be made to the corresponding author.
\bibliographystyle{mnras} 
\bibliography{GridsZ020.bib} 

\appendix

\section{Summary table of the model properties}
Table\,\ref{TabListModels} list the key properties of the present grid of models.

\begin{landscape}
\begin{table}
\caption{Properties of the $Z = 0.020$ stellar models at the end of the H-, He-, and C-burning phases. Columns 1 to 4 give the initial mass of the model, the initial ratio between the rotational velocity at the equator and the critical rotational velocity, the initial rotational velocity at the equator, and the time-averaged equatorial surface velocity during the MS phase, respectively. Columns 5 to 11 present properties of the stellar models at the end of the core H-burning phase: age, total mass, surface equatorial velocity, ratio of the equatorial surface velocity to the critical velocity, mass fraction of helium at the surface, ratios in mass fraction of the nitrogen to carbon abundances at the surface, and of nitrogen to oxygen at the surface,  respectively.  Columns 12 to 17 and 18 to 22 show properties of the models at the end of the core He- and C-burning phases, 
respectively. The columns labeled $M$, $V_{\rm eq}$, $Y_{\rm surf}$, N/C and N/O have the same meaning as the corresponding columns for the end of the core hydrogen burning phase. The quantity $\tau_{\rm He}$ ($\tau_{\rm C}$) corresponds to the duration of the He- (C-) core burning phase.}
\centering
\scalebox{0.66}{\begin{tabular}{rrrr|rrrrrrr|rrrrrr|rrrrr}
\hline\hline
\multicolumn{4}{c|}{} & \multicolumn{7}{c|}{End of H-burning} & \multicolumn{6}{c|}{End of He-burning} & \multicolumn{5}{c}{End of C-burning}\\
 $M_\text{ini}$ & $V_\text{ini}/V_\text{crit}$ & $V_\text{eq}$ & $\bar{V}_\text{MS}$ & $\tau_\text{H}$ & $M$ & $V_\text{eq}$ & $V_\text{eq}/V_\text{crit}$ & $Y_\text{surf}$ & $\text{N}/\text{C}$ & $\text{N}/\text{O}$ & $\tau_\text{He}$ & $M$ & $V_\text{eq}$  & $Y_\text{surf}$ & $\text{N}/\text{C}$ & $\text{N}/\text{O}$ & $\tau_\text{C}$ & $M$ & $Y_\text{surf}$ & $\text{N}/\text{C}$ & $\text{N}/\text{O}$ \\
 $M_{\odot}$ & & \multicolumn{2}{c|}{km s$^{-1}$} & Myr & $M_{\odot}$ & km s$^{-1}$ & & \multicolumn{3}{c|}{mass fract.} & Myr & $M_{\odot}$ & km s$^{-1}$ & \multicolumn{3}{c|}{mass fract.} & kyr & $M_{\odot}$ & \multicolumn{3}{c}{mass fract.}\\
\hline
 $300.00$ & $0.00$ & $  0.$ & $  0.$ & $    2.194$ & $ 49.48$ & $  - $ & $  -  $ & $0.9785$ & $ 65.95$ & $ 94.88$ & $    0.331$ & $ 22.32$ & $  - $ & $0.2366$ & $  0.00$ & $  0.00$ & $    0.015$ & $ 22.23$ & $0.2323$ & $  0.00$ & $  0.00$ \\
 $300.00$ & $0.40$ & $419.$ & $ 35.$ & $    2.286$ & $ 59.16$ & $  1.$ & $0.002$ & $0.9775$ & $ 62.06$ & $ 94.90$ & $    0.319$ & $ 25.34$ & $  1.$ & $0.2355$ & $  0.00$ & $  0.00$ & $    0.013$ & $ 25.24$ & $0.2312$ & $  0.00$ & $  0.00$ \\
 $200.00$ & $0.00$ & $  0.$ & $  0.$ & $    2.349$ & $ 78.23$ & $  - $ & $  -  $ & $0.9205$ & $ 82.92$ & $ 84.84$ & $    0.297$ & $ 35.76$ & $  - $ & $0.2163$ & $  0.00$ & $  0.00$ & $    0.005$ & $ 35.65$ & $0.2125$ & $  0.00$ & $  0.00$ \\
 $200.00$ & $0.40$ & $421.$ & $ 56.$ & $    2.566$ & $ 73.51$ & $  <1$ & $0.000$ & $0.9053$ & $ 82.54$ & $ 83.16$ & $    0.298$ & $ 34.75$ & $  5.$ & $0.2287$ & $  0.00$ & $  0.00$ & $    0.006$ & $ 34.64$ & $0.2248$ & $  0.00$ & $  0.00$ \\
 $150.00$ & $0.00$ & $  0.$ & $  0.$ & $    2.504$ & $ 66.47$ & $  - $ & $  -  $ & $0.8943$ & $ 86.41$ & $ 81.23$ & $    0.307$ & $ 31.01$ & $  - $ & $0.2253$ & $  0.00$ & $  0.00$ & $    0.008$ & $ 30.92$ & $0.2218$ & $  0.00$ & $  0.00$ \\
 $150.00$ & $0.40$ & $408.$ & $ 71.$ & $    2.838$ & $ 60.76$ & $  1.$ & $0.003$ & $0.8873$ & $ 85.22$ & $ 80.22$ & $    0.316$ & $ 25.89$ & $  <1$ & $0.2349$ & $  0.00$ & $  0.00$ & $    0.011$ & $ 25.79$ & $0.2309$ & $  0.00$ & $  0.00$ \\
 $120.00$ & $0.00$ & $  0.$ & $  0.$ & $    2.663$ & $ 56.69$ & $  - $ & $  -  $ & $0.8460$ & $ 91.24$ & $ 76.78$ & $    0.326$ & $ 23.49$ & $  - $ & $0.2235$ & $  0.00$ & $  0.00$ & $    0.016$ & $ 23.40$ & $0.2179$ & $  0.00$ & $  0.00$ \\
 $120.00$ & $0.40$ & $383.$ & $ 82.$ & $    3.108$ & $ 52.67$ & $  2.$ & $0.004$ & $0.8690$ & $ 87.75$ & $ 77.77$ & $    0.328$ & $ 22.35$ & $  <1$ & $0.2380$ & $  0.00$ & $  0.00$ & $    0.017$ & $ 22.26$ & $0.2354$ & $  0.00$ & $  0.00$ \\
 $ 85.00$ & $0.00$ & $  0.$ & $  0.$ & $    3.005$ & $ 45.81$ & $  - $ & $  -  $ & $0.6735$ & $108.01$ & $ 67.75$ & $    0.360$ & $ 16.29$ & $  - $ & $0.2458$ & $  0.00$ & $  0.00$ & $    0.038$ & $ 16.21$ & $0.2405$ & $  0.00$ & $  0.00$ \\
 $ 85.00$ & $0.40$ & $364.$ & $ 96.$ & $    3.628$ & $ 42.72$ & $  4.$ & $0.006$ & $0.8153$ & $ 93.00$ & $ 72.18$ & $    0.356$ & $ 16.73$ & $  <1$ & $0.2536$ & $  0.00$ & $  0.00$ & $    0.038$ & $ 16.64$ & $0.2480$ & $  0.00$ & $  0.00$ \\
 $ 60.00$ & $0.00$ & $  0.$ & $  0.$ & $    3.498$ & $ 33.75$ & $  - $ & $  -  $ & $0.5567$ & $135.85$ & $ 55.75$ & $    0.413$ & $ 10.86$ & $  - $ & $0.2752$ & $  0.00$ & $  0.00$ & $    0.163$ & $ 10.77$ & $0.2681$ & $  0.00$ & $  0.00$ \\
 $ 60.00$ & $0.40$ & $338.$ & $120.$ & $    4.339$ & $ 34.87$ & $  4.$ & $0.007$ & $0.7564$ & $ 81.73$ & $ 42.13$ & $    0.388$ & $ 12.96$ & $ 24.$ & $0.2725$ & $  0.00$ & $  0.00$ & $    0.099$ & $ 12.87$ & $0.2665$ & $  0.00$ & $  0.00$ \\
 $ 40.00$ & $0.00$ & $  0.$ & $  0.$ & $    4.377$ & $ 33.96$ & $  - $ & $  -  $ & $0.2735$ & $  0.29$ & $  0.12$ & $    0.466$ & $ 11.39$ & $  - $ & $0.2138$ & $  0.00$ & $  0.00$ & $    0.140$ & $ 11.33$ & $0.2113$ & $  0.00$ & $  0.00$ \\
 $ 40.00$ & $0.40$ & $302.$ & $139.$ & $    5.562$ & $ 30.60$ & $  5.$ & $0.013$ & $0.5351$ & $ 14.61$ & $  4.67$ & $    0.424$ & $ 11.72$ & $  <1$ & $0.2752$ & $  0.00$ & $  0.00$ & $    0.129$ & $ 11.63$ & $0.2707$ & $  0.00$ & $  0.00$ \\
 $ 32.00$ & $0.00$ & $  0.$ & $  0.$ & $    5.120$ & $ 29.67$ & $  - $ & $  -  $ & $0.2735$ & $  0.29$ & $  0.12$ & $    0.544$ & $ 10.93$ & $  - $ & $0.7779$ & $ 91.58$ & $ 58.69$ & $    0.193$ & $ 10.71$ & $0.9805$ & $ 80.98$ & $ 81.54$ \\
 $ 32.00$ & $0.40$ & $  3.$ & $163.$ & $    6.705$ & $ 26.90$ & $  7.$ & $0.016$ & $0.4779$ & $  7.55$ & $  2.35$ & $    0.504$ & $  9.88$ & $  5.$ & $0.2633$ & $  0.00$ & $  0.00$ & $    0.228$ & $  9.80$ & $0.2551$ & $  0.00$ & $  0.00$ \\
 $ 25.00$ & $0.00$ & $  0.$ & $  0.$ & $    6.228$ & $ 23.98$ & $  - $ & $  -  $ & $0.2735$ & $  0.29$ & $  0.12$ & $    0.702$ & $  8.27$ & $  - $ & $0.5438$ & $166.45$ & $  6.28$ & $    0.483$ & $  8.04$ & $0.9804$ & $ 63.69$ & $ 73.03$ \\
 $ 25.00$ & $0.40$ & $  3.$ & $192.$ & $    8.057$ & $ 23.12$ & $  4.$ & $0.011$ & $0.3752$ & $  3.67$ & $  1.03$ & $    0.598$ & $  9.24$ & $  <1$ & $0.9582$ & $  4.31$ & $ 16.11$ & $    0.292$ & $  9.08$ & $0.9563$ & $  2.25$ & $ 12.86$ \\
 $ 20.00$ & $0.00$ & $  0.$ & $  0.$ & $    7.595$ & $ 19.60$ & $  - $ & $  -  $ & $0.2735$ & $  0.29$ & $  0.12$ & $    0.864$ & $  8.83$ & $  - $ & $0.4950$ & $ 40.42$ & $  3.00$ & $    1.342$ & $  8.45$ & $0.5032$ & $ 56.94$ & $  3.27$ \\
 $ 20.00$ & $0.40$ & $  3.$ & $209.$ & $    9.518$ & $ 19.37$ & $  1.$ & $0.003$ & $0.3093$ & $  2.16$ & $  0.56$ & $    0.799$ & $  7.54$ & $  <1$ & $0.6253$ & $137.80$ & $  3.67$ & $    0.852$ & $  7.27$ & $0.7613$ & $130.19$ & $ 10.91$ \\
 $ 15.00$ & $0.00$ & $  0.$ & $  0.$ & $   10.837$ & $ 14.78$ & $  - $ & $  -  $ & $0.2735$ & $  0.29$ & $  0.12$ & $    1.369$ & $ 13.18$ & $  - $ & $0.3265$ & $  2.30$ & $  0.58$ & $    4.868$ & $ 13.09$ & $0.3505$ & $  3.05$ & $  0.73$ \\
 $ 15.00$ & $0.40$ & $255.$ & $194.$ & $   13.545$ & $ 14.64$ & $140.$ & $0.343$ & $0.2970$ & $  2.08$ & $  0.50$ & $    1.304$ & $ 10.97$ & $  <1$ & $0.3838$ & $  5.65$ & $  0.99$ & $    2.075$ & $ 10.83$ & $0.4013$ & $  6.36$ & $  1.08$ \\
 $ 12.00$ & $0.00$ & $  0.$ & $  0.$ & $   15.151$ & $ 11.93$ & $  - $ & $  -  $ & $0.2735$ & $  0.29$ & $  0.12$ & $    1.994$ & $ 11.61$ & $  - $ & $0.3120$ & $  1.83$ & $  0.49$ & $   10.318$ & $ 11.56$ & $0.3122$ & $  1.98$ & $  0.51$ \\
 $ 12.00$ & $0.40$ & $261.$ & $193.$ & $   18.556$ & $ 11.89$ & $182.$ & $0.452$ & $0.2850$ & $  1.54$ & $  0.39$ & $    1.965$ & $ 10.45$ & $  1.$ & $0.3413$ & $  4.72$ & $  0.79$ & $    4.862$ & $ 10.36$ & $0.3553$ & $  5.26$ & $  0.85$ \\
 \cline{18-22}
 $  9.00$ & $0.00$ & $  0.$ & $  0.$ & $   26.110$ & $  8.99$ & $  - $ & $  -  $ & $0.2735$ & $  0.29$ & $  0.12$ & $    3.486$ & $  8.84$ & $  - $ & $0.2935$ & $  1.58$ & $  0.42$ & \\
 $  9.00$ & $0.40$ & $  2.$ & $182.$ & $   31.679$ & $  8.99$ & $  2.$ & $0.005$ & $0.2791$ & $  1.03$ & $  0.30$ & $    3.413$ & $  8.79$ & $  1.$ & $0.3403$ & $  4.77$ & $  0.78$ & \\
 $  7.00$ & $0.00$ & $  0.$ & $  0.$ & $   42.215$ & $  7.00$ & $  - $ & $  -  $ & $0.2735$ & $  0.29$ & $  0.12$ & $    7.103$ & $  6.93$ & $  - $ & $0.2908$ & $  1.52$ & $  0.40$ & \\
 $  7.00$ & $0.40$ & $230.$ & $175.$ & $   52.260$ & $  7.00$ & $186.$ & $0.501$ & $0.2764$ & $  0.74$ & $  0.24$ & $    6.949$ & $  6.90$ & $  2.$ & $0.3315$ & $  4.07$ & $  0.72$ & \\
 $  5.00$ & $0.00$ & $  0.$ & $  0.$ & $   93.975$ & $  5.00$ & $  - $ & $  -  $ & $0.2735$ & $  0.29$ & $  0.12$ & $   18.718$ & $  4.97$ & $  - $ & $0.2902$ & $  1.46$ & $  0.39$ & \\
 $  5.00$ & $0.40$ & $217.$ & $162.$ & $  114.744$ & $  5.00$ & $165.$ & $0.488$ & $0.2747$ & $  0.51$ & $  0.18$ & $   17.289$ & $  4.95$ & $  3.$ & $0.3279$ & $  3.35$ & $  0.66$ & \\
 $  4.00$ & $0.00$ & $  0.$ & $  0.$ & $  165.064$ & $  4.00$ & $  - $ & $  -  $ & $0.2735$ & $  0.29$ & $  0.12$ & $   44.431$ & $  3.98$ & $  - $ & $0.2930$ & $  1.47$ & $  0.39$ & \\
 $  4.00$ & $0.40$ & $189.$ & $153.$ & $  201.563$ & $  4.00$ & $147.$ & $0.473$ & $0.2743$ & $  0.43$ & $  0.16$ & $   39.392$ & $  3.97$ & $  4.$ & $0.3286$ & $  3.02$ & $  0.63$ & \\
 $  3.00$ & $0.00$ & $  0.$ & $  0.$ & $  354.769$ & $  3.00$ & $  - $ & $  -  $ & $0.2735$ & $  0.29$ & $  0.12$ & $  135.869$ & $  2.99$ & $  - $ & $0.2947$ & $  1.45$ & $  0.38$ & \\
 $  3.00$ & $0.40$ & $193.$ & $142.$ & $  441.187$ & $  3.00$ & $131.$ & $0.460$ & $0.2740$ & $  0.38$ & $  0.14$ & $  115.809$ & $  2.98$ & $  5.$ & $0.3308$ & $  2.79$ & $  0.60$ & \\
 $  2.50$ & $0.00$ & $  0.$ & $  0.$ & $  589.831$ & $  2.50$ & $  - $ & $  -  $ & $0.2735$ & $  0.29$ & $  0.12$ & $  241.474$ & $  2.49$ & $  - $ & $0.2920$ & $  1.36$ & $  0.35$ & \\
 $  2.50$ & $0.40$ & $187.$ & $136.$ & $  742.734$ & $  2.50$ & $122.$ & $0.450$ & $0.2739$ & $  0.36$ & $  0.14$ & $  190.089$ & $  2.49$ & $  5.$ & $0.3290$ & $  2.60$ & $  0.55$ & \\
\cline{12-17}
 $  2.00$ & $0.00$ & $  0.$ & $  0.$ & $ 1119.481$ & $  2.00$ & $  - $ & $  -  $ & $0.2735$ & $  0.29$ & $  0.12$ & \\
 $  2.00$ & $0.40$ & $185.$ & $133.$ & $ 1436.772$ & $  2.00$ & $118.$ & $0.456$ & $0.2740$ & $  0.34$ & $  0.13$ & \\
 $  1.70$ & $0.00$ & $  0.$ & $  0.$ & $ 1826.804$ & $  1.70$ & $  - $ & $  -  $ & $0.2735$ & $  0.29$ & $  0.12$ & \\
 $  1.70$ & $0.40$ & $150.$ & $127.$ & $ 2358.293$ & $  1.70$ & $109.$ & $0.422$ & $0.2741$ & $  0.33$ & $  0.13$ & \\
 $  1.50$ & $0.00$ & $  0.$ & $  0.$ & $ 2543.919$ & $  1.50$ & $  - $ & $  -  $ & $0.2735$ & $  0.29$ & $  0.12$ & \\
 $  1.50$ & $0.40$ & $150.$ & $ 10.$ & $ 3024.281$ & $  1.50$ & $  9.$ & $0.000$ & $0.2747$ & $  0.36$ & $  0.13$ & \\
 $  1.35$ & $0.00$ & $  0.$ & $  0.$ & $ 3743.685$ & $  1.35$ & $  - $ & $  -  $ & $0.2735$ & $  0.29$ & $  0.12$ & \\
 $  1.35$ & $0.40$ & $ 28.$ & $  7.$ & $ 4122.357$ & $  1.35$ & $  6.$ & $0.000$ & $0.2744$ & $  0.31$ & $  0.12$ & \\
 $  1.25$ & $0.00$ & $  0.$ & $  0.$ & $ 5025.342$ & $  1.25$ & $  - $ & $  -  $ & $0.2285$ & $  0.29$ & $  0.11$ & \\
 $  1.25$ & $0.40$ & $ 26.$ & $  5.$ & $ 5343.693$ & $  1.25$ & $  4.$ & $0.000$ & $0.2623$ & $  0.30$ & $  0.12$ & \\
 $  1.10$ & $0.00$ & $  0.$ & $  0.$ & $ 6238.572$ & $  1.10$ & $  - $ & $  -  $ & $0.2207$ & $  0.29$ & $  0.11$ & \\
 $  1.10$ & $0.40$ & $ 50.$ & $  3.$ & $ 6339.092$ & $  1.10$ & $  3.$ & $0.000$ & $0.2581$ & $  0.29$ & $  0.12$ & \\
 $  1.00$ & $0.00$ & $  0.$ & $  0.$ & $ 9677.035$ & $  1.00$ & $  - $ & $  -  $ & $0.2202$ & $  0.29$ & $  0.11$ & \\
 $  1.00$ & $0.40$ & $ 50.$ & $  2.$ & $ 9895.752$ & $  1.00$ & $  2.$ & $0.000$ & $0.2541$ & $  0.29$ & $  0.11$ & \\
 $  0.90$ & $0.00$ & $  0.$ & $  0.$ & $15196.086$ & $  0.90$ & $  - $ & $  -  $ & $0.2129$ & $  0.29$ & $  0.11$ & \\
 $  0.90$ & $0.40$ & $ 12.$ & $  2.$ & $15600.981$ & $  0.90$ & $  1.$ & $0.000$ & $0.2481$ & $  0.29$ & $  0.11$ & \\
 $  0.80$ & $0.00$ & $  0.$ & $  0.$ & $24110.325$ & $  0.80$ & $  - $ & $  -  $ & $0.2011$ & $  0.29$ & $  0.11$ & \\
 $  0.80$ & $0.40$ & $  6.$ & $  1.$ & $24858.578$ & $  0.80$ & $  1.$ & $0.000$ & $0.2391$ & $  0.29$ & $  0.11$ & \\
\cline{1-11}
\end{tabular}
}
\label{TabListModels}
\end{table}
\end{landscape}

\label{lastpage}
\end{document}